\pdfpageattr{/Group << /S /Transparency /I true /CS /DeviceRGB>>}
\documentclass[aps,prc,reprint,superscriptaddress,showpacs,amsmath,nofootinbib]{revtex4-1}
\usepackage[T1]{fontenc}
\usepackage[utf8]{inputenc}

\usepackage{todonotes}
\usepackage{isotope}
\usepackage{xcolor}
\usepackage{graphicx}
\usepackage[load-configurations=abbreviations]{siunitx}
\usepackage{braket}
\usepackage{mathtools}
\usepackage{ifpdf}
\usepackage{booktabs}

\newcommand{\doubletoprule}{\toprule[\lightrulewidth]\toprule[\lightrulewidth]}
\newcommand{\doublebottomrule}{\bottomrule[\lightrulewidth]\bottomrule[\lightrulewidth]}

\usepackage{txfonts}

\usepackage{tikz}
\usetikzlibrary{calc}

\ifpdf\else\renewcommand{\href}[2]{#2}\fi

\usepackage{hyperref}
\usepackage[capitalize]{cleveref}

\crefname{section}{Sect.}{Sects.}

\crefname{equation}{}{}
\Crefname{equation}{Equation}{Equations}

\makeatletter
\DeclareUnicodeCharacter{014C}{\@tabacckludge=O} % LATIN CAPITAL LETTER O WITH MACRON
\DeclareUnicodeCharacter{014D}{\@tabacckludge=o} % LATIN SMALL LETTER O WITH MACRON
\makeatother

\DeclareSIUnit{\GeV}{\giga\electronvolt}
\DeclareSIUnit{\MeV}{\mega\electronvolt}
\DeclareSIUnit{\keV}{\kilo\electronvolt}
\DeclareSIUnit\MeVc{\MeV\per\text{\ensuremath{c}}}
\DeclareSIUnit{\fm}{\femto\meter}

\newcommand{\partd}[2]{\frac{\partial #1}{\partial #2}}
\newcommand{\comm}[2]{\lbrack #1,#2\rbrack}
\newcommand{\kronecker}[2]{\delta^{#1}_{#2}}
\newcommand{\clebsch}[6]{\ensuremath{C_{#1#2,#3#4}^{#5#6}}} % Varshalovich convention
\newcommand{\sixj}[6]{\begingroup\setlength{\arraycolsep}{0.2em}\begin{Bmatrix} #1 & #2 & #3 \\ #4 & #5 & #6 \end{Bmatrix}\endgroup}
\newcommand{\ninej}[9]{\begingroup\setlength{\arraycolsep}{0.2em}\begin{Bmatrix} #1 & #2 & #3 \\ #4 & #5 & #6 \\ #7 & #8 & #9 \end{Bmatrix}\endgroup}
\newcommand{\hob}[6]{\langle\!\langle #1,#2\mspace{4mu}|\mspace{4mu}#3,#4:#5\rangle\!\rangle_{#6}}

\DeclareMathOperator{\sgn}{sgn}

\newcommand{\abra}[1]{\prescript{}{a}{\bra{#1}}}
\newcommand{\keta}[1]{\ket{#1}_a}

\newcommand{\abraketa}[1]{\prescript{}{a}{\braket{#1}}\mathclose{\vphantom{\braket{#1}}}_{a}}

\newcommand{\pcl}{\chi}
\newcommand{\pw}{\nu}
\newcommand{\Strng}{\mathcal{S}}
\newcommand{\jhat}{\hat{\jmath}}
\newcommand{\Nmax}{\ensuremath{N_\text{max}}}
\newcommand{\modelspace}{\mathcal{M}}

\newcommand*{\op}[1]{{#1}}
\newcommand{\Ham}{\op{H}}
\newcommand{\Tint}{\op{T}_{\text{int}}}
\newcommand{\Valpha}[1][]{\op{V}_{\alpha #1}}
\newcommand{\Mass}{\op{M}}
\newcommand{\Antisym}{\op{\mathcal{A}}}
\newcommand{\Perm}{\op{\mathcal{P}}}

\newcommand*{\vect}[1]{\vec{#1}}

\newcommand{\pheq}{\phantom{{}={}}}

\newcommand{\Ncal}{\mathcal{N}}
\newcommand{\Lcal}{\mathcal{L}}
\newcommand{\Jcal}{\mathcal{J}}

\newcommand{\Jred}[1]{\bar{#1}}

\newcommand{\Am}[1]{A\text{-}#1}
\newcommand{\NNN}{\ensuremath{NNN}}
\newcommand{\NN}{\ensuremath{NN}}
\newcommand{\YN}{\ensuremath{YN}}
\newcommand{\YNN}{\ensuremath{YNN}}
\newcommand{\YNNN}{\ensuremath{YNNN}}
\newcommand{\NY}{\YN}

\newcommand{\submass}[2]{\ensuremath{M_{#1,#2}}}

\allowdisplaybreaks[1]

\begin{document}

\title{Hypernuclear No-Core Shell Model}

\author{Roland Wirth}
\email{roland.wirth@physik.tu-darmstadt.de}
\affiliation{Institut f\"ur Kernphysik -- Theoriezentrum, TU Darmstadt, Schlossgartenstr. 2, 64289 Darmstadt, Germany}

\author{Daniel Gazda}
\affiliation{Department of Physics, Chalmers University of Technology, SE-412 96 G\"oteborg, Sweden}
\affiliation{Nuclear Physics Institute, 25068 \v{R}e\v{z}, Czech Republic}

\author{Petr Navr\'atil}
\affiliation{TRIUMF, 4004 Wesbrook Mall, Vancouver, British Columbia, V6T 2A3, Canada}

\author{Robert Roth}
\email{robert.roth@physik.tu-darmstadt.de}
\affiliation{Institut f\"ur Kernphysik -- Theoriezentrum, TU Darmstadt, Schlossgartenstr. 2, 64289 Darmstadt, Germany}

\date{\today}

\begin{abstract}%
We extend the No-Core Shell Model (NCSM) methodology to incorporate strangeness degrees of freedom and apply it to single-$\Lambda$ hypernuclei.
After discussing the transformation of the hyperon-nucleon (YN) interaction into Harmonic-Oscillator (HO) basis and the Similarity Renormalization Group transformation applied to it to improve model-space convergence, we present two complementary formulations of the NCSM, one that uses relative Jacobi coordinates and symmetry-adapted basis states to fully exploit the symmetries of the hypernuclear Hamiltonian, and one working in a Slater determinant basis of HO states where antisymmetrization and computation of matrix elements is simple and to which an importance-truncation scheme can be applied.
For the Jacobi-coordinate formulation, we give an iterative procedure for the construction of the antisymmetric basis for arbitrary particle number and present the formulae used to embed two- and three-baryon interactions into the many-body space.
For the Slater-determinant formulation, we discuss the conversion of the YN interaction matrix elements from relative to single-particle coordinates, the importance-truncation scheme that tailors the model space to the description of the low-lying spectrum, and the role of the redundant center-of-mass degrees of freedom.
We conclude with a validation of both formulations in the four-body system, giving converged ground-state energies for a chiral Hamiltonian, and present a short survey of the $A\le7$ hyper-helium isotopes.
\end{abstract}


\maketitle

Strangeness impacts many fields of physics from heavy-ion collisions to nuclear and neutron star structure.
Of particular interest are hypernuclei, which can be produced and studied in the laboratory.
Hypernuclei are many-body systems consisting of nucleons and hyperons, baryons that carry strangeness, like the $\Lambda^0$, $\Sigma^{0,\pm}$, or the $\Xi^{0,-}$.
These hyperons are distinguishable from the nucleons and can be used as probes for the interior structure of the nucleonic core.
Furthermore, hypernuclei extend the isospin SU(2), which is a good approximate symmetry in nuclei, to flavor SU(3) that is broken by the significant mass difference between the strange and the up and down quarks \cite{Gal2016}.
This breaking allows new types of baryon-baryon interactions such as antisymmetric spin-orbit forces, which are forbidden by isospin symmetry \cite{Haidenbauer2013}.

A variety of experiments were performed to study properties of hypernuclei.
From the early emulsion experiments (see, e.g., Ref.~\cite{Davis2005}) to modern accelerator-based experiments, a lot of effort went into the measurement of not only ground-state properties \cite{Miyoshi2003,Cusanno2009,Agnello2012,Nakamura2013,Esser2015,Gogami2016}, but also the determination of hypernuclear spectra by gamma-ray spectroscopy \cite{May1983,Tamura2000,Akikawa2002,Hashimoto2006,Yamamoto2015}.
Even transition strengths are experimentally accessible \cite{Tanida2001}.
This effort was complemented by various theory developments, e.g., Skyrme- and Brueckner-Hartree-Fock models \cite{Halderson1993,Cugnon2000,Vidana2001}, the shell model \cite{Gal1971,*Gal1972,*Gal1978,Millener2001,Millener2005,Millener2008,Millener2010,Millener2012}, cluster models \cite{Motoba1983,Hiyama2001,Hiyama2009,Hiyama2012}, and few-body methods \cite{Miyagawa1995,Nemura2002,Nogga2002,Nogga2005,Nogga2013}.
Recently, Quantum Monte-Carlo methods were developed that make ground-state properties of a wide range of hypernuclei accessible \cite{Lonardoni2013,Lonardoni2014,Lonardoni2015}.

What was missing from this wealth of theoretical approaches was a method that provides systematically improvable calculations for both ground- and excited-state properties of hypernuclei and is flexible in the choice of interactions (unlike the Quantum Monte-Carlo methods which require local ones).
Such an \emph{ab initio} many-body method is the cornerstone for a description of $p$\nobreakdash-shell hypernuclei with interactions derived from chiral effective field theory framework, which is rooted in the symmetries of quantum chromodynamics.
This was the original motivation for the development of the hypernuclear shell model, which we presented recently \cite{Wirth2014}.

In this paper, we describe the steps needed to perform No-Core Shell-Model (NCSM) calculations for single\nobreakdash-$\Lambda$ hypernuclei.
We start with the preparation of the harmonic-oscillator (HO) matrix elements of the hyperon-nucleon (\YN{}) interaction, which are often provided in terms of spin-isospin operators or---in our case---as momentum-space matrix elements (\cref{sec:hamil}).
These matrix elements are optionally subject to a Similarity Renormalization Group (SRG) transformation, and converted to HO representation.
We then present two complementary formulations of the many-body method, one using a basis with good angular momentum and isospin quantum numbers employing translation-invariant Jacobi coordinates (\cref{sec:jncsm}) and one using a Slater determinant basis of HO single-particle states (\cref{sec:itncsm}).
The Jacobi-coordinate formulation (J-NCSM) takes full advantage of the symmetries of the Hamiltonian, drastically reducing the number of basis states in the model space, at the cost of an antisymmetrization procedure that grows exceedingly difficult with the number of particles.
The Slater determinant basis provides trivial antisymmetrization, but the basis states exhibit less of the symmetries and contain the center-of-mass coordinate as redundant degree of freedom.
Due to its simplicity, however, it is easy to implement an importance truncation scheme tailoring the basis to the description of a few target eigenstates, leading to the Importance-Truncated No-Core Shell Model (IT-NCSM).
We conclude with a validation of the J-NCSM and the IT-NCSM in the four-body system (\cref{sec:validation}), where we present absolute energies of the ground and first excited states of \isotope[4][\Lambda]{H} and \isotope[4][\Lambda]{He} for a state-of-the-art chiral hypernuclear Hamiltonian and discuss effects of the SRG transformation, and provide a survey of the hyper-helium chain (\cref{sec:hyperhelium}).

\section{\label{sec:hamil}Hypernuclear Hamiltonian}
\subsection{Hamiltonian and Jacobi Coordinates}
The starting point of the \emph{ab initio} NCSM calculation is the intrinsic Hamiltonian for a system of $A$ nonrelativistic nucleons and hyperons interacting by realistic two-body \NN{}, \YN{}, and three-nucleon (\NNN{}) interactions:
\begin{equation} \label{eq:hamil:hsp}
  \Ham_{\text{int}} = \Delta\Mass + \Tint + \op{V}^{[2]}_{\NN} + \op{V}^{[2]}_{\YN} + \op{V}^{[3]}_{\NNN},
\end{equation}
with intrinsic kinetic energy $\Tint = \op{T} - \op{T}_{\text{c.m.}}$ and a mass term
\begin{equation}
  \Delta\Mass = \sum_i \op{m}_i - M_0
\end{equation}
accounting for the rest-mass difference of the $\Lambda$ and $\Sigma$ hyperons.
Here, $M_0$ is the reference mass given by the total rest mass of the protons, neutrons, and $\Lambda$ hyperons pertinent to the system under consideration.
We use physical particle masses for the IT-NCSM; due to isospin coupling the J-NCSM takes isospin-averaged particle masses.
The mass term is necessary because the \YN{} interaction couples $\Lambda N$ and $\Sigma N$ states ($\Lambda N$-$\Sigma N$ conversion), and we have to consider the full coupled-channel system.

Since the intrinsic Hamiltonian \eqref{eq:hamil:hsp} is translation-invariant, it is convenient to introduce relative Jacobi coordinates for the baryons and define basis states with respect to these coordinates.
The transformation between the single-particle (s.p.) and Jacobi coordinates is in general not orthogonal, but an orthogonal transformation is needed to transform HO states between the two coordinate systems.
To address that, we define scaled versions of the s.p.\ coordinates
\begin{equation}
  \vect{x}_i = \sqrt{\frac{m_i}{m_N}}\vect{r}_i,
\end{equation}
where $\vect{r}_i$ and $m_i$ are the coordinate and rest mass of baryon $i$.
The nucleon mass $m_N$ is used as a common scale and the description of the system in terms of the scaled coordinates is the same as in terms of the unscaled ones when all other length scales, such as the oscillator lengths, are scaled accordingly.

There are several different sets of Jacobi coordinates, one of which is given by
\begin{subequations}\label{eq:hamil:jxi}
\begin{align}
\vect{\xi}_0 &= \frac{1}{\sqrt{\submass{1}{A}}}\sum_{i=1}^A \sqrt{m_i} \vect{x}_i,\\
\vect{\xi}_i &=
  \sqrt{\frac{\submass{1}{i} m_{i+1}}{\submass{1}{i+1}}}
  \left(\frac{1}{\submass{1}{i}}\sum_{j=1}^{i}\sqrt{m_j}\vect{x}_j
    -\frac{1}{\sqrt{m_{i+1}}}\vect{x}_{i+1}\right),
\shortintertext{with}
\submass{i}{j}&=\sum_{k=i}^{j}m_k.
\end{align}
\end{subequations}
Analogous transformations are defined for momenta $\vect{p}_i$.
In general, Jacobi coordinates are proportional to differences of center-of-mass (c.m.) coordinates of individual particle subclusters and $\vect{\xi}_0$ is proportional to the c.m.\ coordinate of the $A$\nobreakdash-baryon system.

For the two-body system, the Jacobi coordinates \eqref{eq:hamil:jxi} reduce to
\begin{subequations}\label{eq:hamil:2b jacobi}
\begin{align}
  \vect{\xi}_0 &= \frac1{\sqrt{\submass{1}{2}}}(\sqrt{m_1} \vect{x}_1 + \sqrt{m_2} \vect{x}_2) = \sqrt{\frac{\submass{1}{2}}{m_N}} \vect{R} \label{eq:hamil:2b jacobi cm}\\
  \vect{\xi}_1 &= \sqrt{\frac{m_1m_2}{\submass{1}{2}}}\left(\frac1{\sqrt{m_1}} \vect{x}_1 - \frac1{\sqrt{m_2}} \vect{x}_2\right) = \sqrt{\frac{\mu_{12}}{m_N}} \vect{r} \label{eq:hamil:2b jacobi rel}
\end{align}
\end{subequations}
with total and reduced mass $\submass{1}{2}$ and $\mu_{12}$, respectively.
In this case, the Jacobi coordinates are proportional to the c.m.\ and relative coordinates $\vect{R} = (m_1\vect{r}_1 + m_2\vect{r}_2)/\submass{1}{2}$ and $\vect{r} = \vect{r}_1 - \vect{r}_2$.

Due to translation invariance, the two-body interaction is independent of $\vect{\xi}_0$ and can be represented as a matrix in a partial-wave decomposed relative-momentum basis $\ket{q\pw}$, with relative momentum
\begin{equation}
  \vect{q} = \frac1{\submass{1}{2}}\bigl(m_2 \vect{p}_1 - m_1 \vect{p}_2)
\end{equation}
and $\pw=\{(LS)JM,\pcl_a\pcl_b\}$ collecting the relative orbital angular momentum, coupled spin, total angular momentum (with projection $M$) and isospin quantum numbers of the two particles.
The parentheses denote angular-momentum coupling.
The $\pcl_i=\{\Strng_is_it_im_{t,i}\}$ denote the species (strangeness, spin, isospin and isospin projection) of the involved particles, i.e., proton, neutron, $\Lambda$ and $\Sigma^{0,\pm}$ hyperons for singly-strange hypernuclei.
In case of multi-strange systems the $\Xi^{0,-}$ doublet has to be included as well.

In order to use such a two-body interaction in an NCSM calculation it has to be converted to a HO basis and we optionally subject it to a Similarity Renormalization Group (SRG) transformation to improve convergence of the calculation with respect to model-space size.
We give the details of these steps in the following sections.

\subsection{\label{sec:hamil:relative ho basis}Hyperon-Nucleon Interaction in Relative HO Basis}
A necessary step is the conversion of the relative momen\-tum-space matrix elements to a relative HO basis $\keta{n\pw}$ with radial quantum number $n$, either before or after the SRG evolution.
The subscript denotes a state that is antisymmetric under particle exchange.
Using antisymmetric two-body \YN{} states is convenient in the IT-NCSM because it allows us to treat the hyperon and nucleons identically; it is more efficient to keep the hyperon separate in the J-NCSM, so we work with nonantisymmetric states there.
The formulae for the transformation and SRG evolution are the same in both cases.

The relative two-body basis states are parametrized by the oscillator frequency $\Omega$ that defines the stiffness of the harmonic potential and is connected to the oscillator length
\begin{equation}\label{eq:hamil:oscillator length}
 a_{\pcl_a\pcl_b} = \frac{1}{\sqrt{\mu_{\pcl_a\pcl_b}\Omega}},
\end{equation}
the intrinsic length scale of the oscillator, which depends on the reduced mass $\mu_{\pcl_a\pcl_b}$ of the particles involved.%
\footnote{We use natural units, setting $\hbar=c=1$.}
In a truncated model space, observables in general depend on the basis parameter.
The dependence becomes weaker with increasing model-space size and vanishes for the full Hilbert space.

Before doing the actual transformation, we consider the antisymmetrization of a relative two-body state.
For this system, the antisymmetrization operator reads $\Antisym = \tfrac12 (1 - \Perm)$, where the operator $\Perm$ transposes the particles.
The action of $\Perm$ on a relative HO state is
\begin{align}
 &\Perm \ket{n\pw} = (-1)^{L+S-1}\ket{n\pw[a\leftrightarrow b]},
 \label{eq:hamil:p12 ho state}
\end{align}
where we assumed that the particles have spin $1/2$ and used the symmetry properties of the Clebsch-Gordan coefficients and the equivalence of a particle exchange in coordinate space to a parity operation.
This parity operation results in a phase factor $(-1)^L$ from the spherical harmonics governing the angular dependence of the coordinate space wave function.
The notation $\ket{n\pw[a\leftrightarrow b]}$ denotes a transposition of the particle-species indices $\pcl_a\leftrightarrow\pcl_b$.
The radial part depends only on the absolute value of the relative distance and hence remains unchanged.

Using \cref{eq:hamil:p12 ho state}, we apply the antisymmetrization operator to a HO state, yielding
\begin{align}
  \Antisym\ket{n\pw} &= \tfrac12 \bigl(\ket{n\pw} + (-1)^{L+S}\ket{n\pw[a\leftrightarrow b]}\bigr).
  \label{eq:hamil:antisymmetrized relative state}
\end{align}
The antisymmetrization of relative-momentum states is achie\-ved in exactly the same way because these states differ only in their radial component.

The transformation from relative-momentum basis to HO basis is achieved by expanding the HO state
\begin{align}
 \keta{n\pw} &= \int \mathrm{d}q\,q^2 \keta{q\pw}\abraketa{q\pw|n\pw} = \int \mathrm{d}q\,q^2  \phi^{\pw}_n(q) \keta{q\pw}.
\end{align}
where we eliminated the sum over partial waves due to orthogonality.
The overlap between the two antisymmetrized states is the momentum-space wave function of the HO state
\begin{align}
 \phi^{\pw}_n(q)= \sqrt{\frac{2 a^3_{\pcl_a\pcl_b} n!}{\Gamma(n+L+\tfrac32)}} (-1)^n \mathrm{e}^{-\frac12 \varrho^2} \varrho^L L_n^{(L+\frac12)}\Bigl(\varrho^2\Bigr)
\end{align}
where $\varrho=a_{\pcl_a\pcl_b}q$ is the dimensionless relative momentum, $\Gamma(x)$ is Euler's gamma function and $L_n^{(\alpha)}(x)$ is an associated Laguerre polynomial.

\subsection{\label{sec:hamil:srg}Similarity Renormalization Group Transformation}
Baryon-baryon interactions induce significant short-range correlations due to their repulsive core and strong tensor forces, which manifest in large off-diagonal interaction matrix elements between states with low and high relative momentum.
In order to improve convergence in the model-space sizes reachable with the NCSM, these correlations have to be controlled.

We eliminate short-ranged correlations by evolving the Hamiltonian using SRG transformations \cite{Wegner1994,Wegner2000,Bogner2007}, defined by the flow equation
\begin{equation}\label{eq:hamil:flow equation}
 \partd{\Ham_\alpha}{\alpha}=\comm{\op{\eta}_\alpha}{\Ham_\alpha},
\end{equation}
where $\Ham_\alpha$ is the evolved Hamiltonian, $\op{\eta}_\alpha$ is the generator of the transformation and $\alpha$ denotes the continuous flow parameter controlling the evolution.
The anti-Hermitian generator $\op{\eta}_\alpha$ may be chosen freely to achieve a desired behavior of the transformation.
We adopt one of the most common choices~\cite{Bogner2007}, namely
\begin{equation}\label{eq:hamil:generator}
 \op{\eta}_\alpha = (2\mu_N)^2\comm{\Tint}{\Ham_\alpha},
\end{equation}
where the reduced mass in the \NN{} system $\mu_N=m_N/2$ is used solely to set the unit and scale of the flow parameter.
The operator $\Tint$ is the intrinsic kinetic energy (see \cref{app:tint and com} for a general expression).
The evolution governed by this generator has a fixpoint when the evolved Hamiltonian $\Ham_\alpha$ commutes with the intrinsic kinetic energy.
In that case, the Hamiltonian is diagonal in momentum-space representation and states with different momenta are decoupled completely.

By partitioning the Hamiltonian $\Ham_\alpha$ such that only the interaction $\Valpha$ depends on the flow parameter,
\begin{equation}
 \Ham_\alpha = \Mass + \Tint + \Valpha,
\end{equation}
the flow equation \eqref{eq:hamil:flow equation} for the generator \eqref{eq:hamil:generator} simplifies to
\begin{align}
  \partd{\Valpha}{\alpha} = (2\mu_N)^{2}\comm{\comm{\Tint}{\Valpha}}{\Mass + \Tint + \Valpha}.\label{eq:hamil:flow equation tint}
\end{align}
Note that even if at the start of the evolution $\Valpha[=0]$ is a two-body interaction, the commutator contains up to four-body terms and the evolution to finite flow parameter values induces interactions of higher particle rank.
The evolved Hamiltonian $\Ham_\alpha$ of an $A$\nobreakdash-body system is thus a genuine $A$\nobreakdash-body operator.

The treatment of the full $A$\nobreakdash-body Hamiltonian is not feasible and the importance of the induced many-body interactions, given a proper generator choice and a sufficiently small flow parameter, is rapidly declining with increasing particle rank.
We perform a truncation of the nucleon--nucleon and three-nucleon interactions at the three-body level using the methods described in \cite{Jurgenson2009,Roth2011,Roth2014}.
Here, we focus on the evolution of the combined two-nucleon and hyperon--nucleon interaction at the two-body level.

\begin{figure}
  \centering%
  \includegraphics{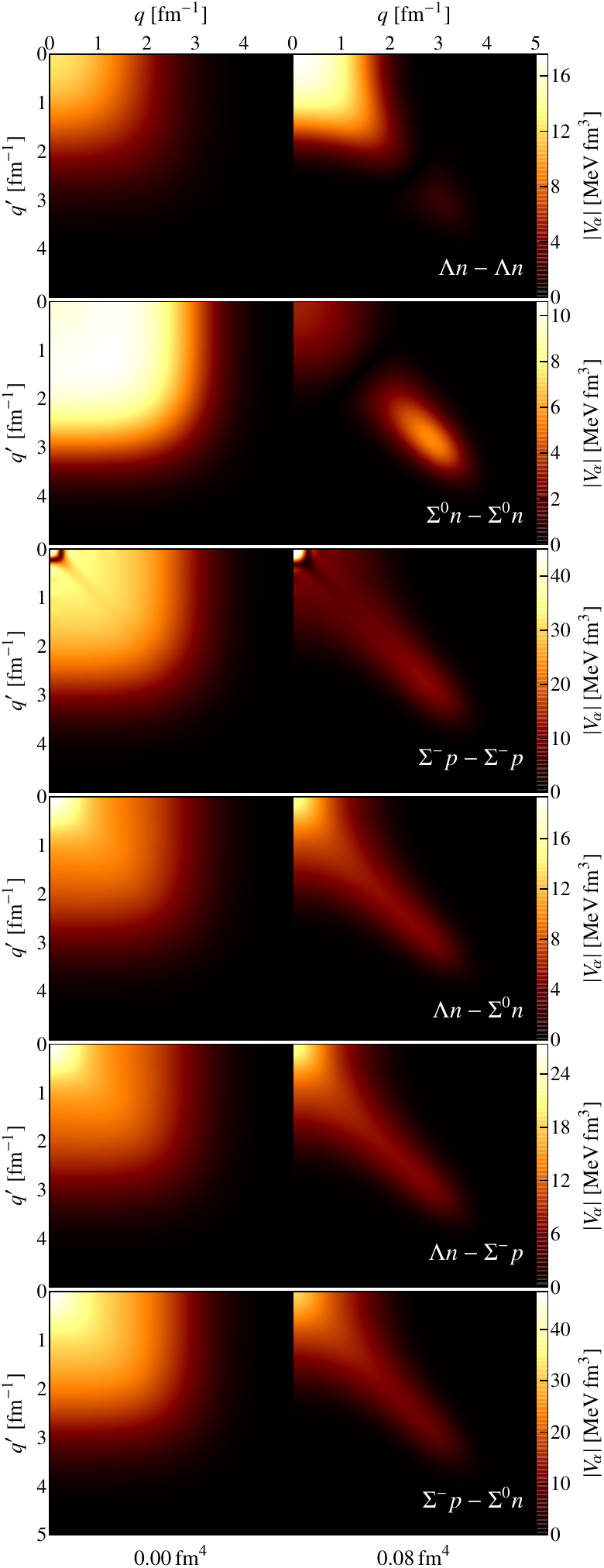}%
  \caption{\label{fig:hamil:matrixelements}%
    (Color online) Momentum-space matrix elements of the $Q=0$, $\Strng=-1$ $^1\mathrm{S}_0$ partial wave for the bare and evolved LO YN interaction \cite{Polinder2006} with a regulator cutoff of $\Lambda=\SI{600}{\MeVc}$.
    The particle-combinations are coupled by the transition matrix elements in the bottom three rows and have to be evolved simultaneously.
    The low-momentum matrix elements of the $\Sigma^-p$ channel (third row) are dominated by the Coulomb interaction.
  }
\end{figure}

In order to solve the flow equation \eqref{eq:hamil:flow equation tint}, we need to convert this operator equation into a system of coupled ordinary differential equations.
We do this by evaluating \cref{eq:hamil:flow equation tint} on a sufficiently large set of basis states.
Since the interaction matrix elements are initially given in momentum-space and $\Tint$ is diagonal, an obvious choice is to use relative-momentum states as basis.
An alternative is the HO basis.
Both approaches are equivalent if the respective model spaces are sufficiently large.

First, we show the evolution of the Hamiltonian in relative momentum-space representation.
In this basis, $\Tint$ and $\Mass$ are diagonal and the respective eigenvalue relations can be used.
The SRG flow equation in momentum space is an integro-differential equation which we turn into an ordinary matrix differential equation by discretizing the relative-momentum basis.

To get the flow equation in terms of matrix elements of $\Valpha$, we introduce discretized identity operators
\begin{equation}
 \op{1} = \sum_\pw \int \mathrm{d}q\, q^2 \keta{q\pw}\abra{q\pw} \approx \sum_\pw\sum^{q_\text{max}}_{q}\Delta q\,q^2 \keta{q\pw}\abra{q\pw} \label{eq:hamil:discretized identity}
\end{equation}
in \cref{eq:hamil:flow equation tint} after expanding the commutators.
Employing the notation
\begin{equation}
 V_\alpha^{\pw,\pw'}(q,q') = \abraketa{q\pw|\Valpha|q'\pw'},
\end{equation}
the flow equation \cref{eq:hamil:flow equation tint} becomes
\begin{align}
  \partd{}{\alpha}V_\alpha^{\pw,\pw'}(q,q') &= -\Bigl(\tfrac{\mu_N}{\mu}q^2 - \tfrac{\mu_N}{\mu'}q'^2\Bigr)^2 V_\alpha^{\pw,\pw'}(q,q') \notag\\
  &\pheq - 2\mu_N \Bigl(M-M'\Bigr)\Bigl(\tfrac{\mu_N}{\mu}q^2 - \tfrac{\mu_N}{\mu'}q'^2\Bigr) V_\alpha^{\pw,\pw'}(q,q') \notag\\
  &\pheq + 2\mu_N \smashoperator{\sum_{q''\pw''}} \Delta q''\, q''^2 \Bigl(\tfrac{\mu_N}{\mu} q^2 + \tfrac{\mu_N}{\mu'} q'^2 - 2 \tfrac{\mu_N}{\mu''} q''^2\Bigr) \notag\\
  &\hphantom{{}+ 2\mu_N \sum_{q''\pw''}} {}\times V_\alpha^{\pw,\pw''}(q,q'')V_\alpha^{\pw'',\pw'}(q'',q').\label{eq:hamil:matrix flow equation}
\end{align}
Here, $M,\mu$ ($M',\mu'$) denote the total and reduced masses of the particles in $\pw$ ($\pw'$); $\mu''$ is the reduced mass of the intermediate-state particles that are summed over.
The second line of the flow equation is due to the mass term in the Hamiltonian, and is not present for nucleons.
The other terms get factors of $\mu_N/\mu$ compared to the nucleonic SRG evolution.
Note that the last term couples not only matrix elements with different relative momenta, but also different partial waves.

We thus have to perform a simultaneous evolution of all partial waves that are connected via the potential $\Valpha$:
Partial waves with $L,L'=J\pm1$ and $S,S'=1$ due to tensor forces, with $L,L'=J$ and $S,S'=0,1$ due to anti-symmetric spin-orbit forces, and with same total charge $Q$ and strangeness $\Strng$ due to $\Lambda N$-$\Sigma N$ conversion.
Tensor forces and antisymmetric spin-orbit forces couple disjoint sets of partial waves so that we have a simultaneous evolution of either $\pw,\pw'=\{(L=J\pm1,S=1)JM,\pcl_a\pcl_b\}$ or $\pw,\pw'=\{(L=J,S=\{0,1\})JM,\pcl_a\pcl_b\}$ for all particle combinations $\pcl_a\pcl_b$ with the same $Q$ and $\Strng$.
For the $\Strng=-1$ case, where we have up to three particle combinations, this leads to the simultaneous evolution of up to six partial waves.
Using this information, we employ a standard Runge-Kutta-Fehlberg \cite{Fehlberg1970} solver to evolve \cref{eq:hamil:matrix flow equation} with initial condition $\Valpha[=0] = \op{V}$ up to a given flow parameter $\alpha$.
As an example, the effect of the transformation on the $Q=0$, $\Strng=-1$, $^1\mathrm{S}_0$ partial-wave matrix elements is shown in \cref{fig:hamil:matrixelements}.

The evolution in HO basis is simpler because the basis is discrete from the outset and the resolution of the identity does not contain any additional factors.
After choosing a maximum radial quantum number we calculate matrix representations of $\Ham$ and $\Tint$, considering the full coupled-channel problem.
Then, we solve the initial value problem numerically, evaluating \cref{eq:hamil:flow equation} by computing the double matrix commutator.
After the evolution we recover the interaction matrix elements by subtracting the unevolved matrix representations of $\Tint$ and $\Mass$.

\section{\label{sec:jncsm}Jacobi-Coordinate Formulation}
The SRG-evolved two- and three-body interactions can be used in any basis-expansion-based many-body approach to compute properties of hypernuclei.
In the following, we present a particularly powerful \emph{ab initio} method that is suitable for light hypernuclei: the No-Core Shell Model (NCSM).
We can calculate not only ground- and excited-state energies but the many-body wave functions themselves in this framework.
This enables us to access observables beyond energies.

The NCSM is based on an expansion of the total wave function in a many-body HO basis.
It can be formulated equivalently in terms of single-particle (cf.\ \cref{sec:itncsm}) and Jacobi coordinates.
The Jacobi-coordinate formulation uses the symmetries of the intrinsic Hamiltonian to omit the c.m.\ coordinate $\vect{\xi}_0$ and build a $JT$\nobreakdash-coupled HO basis depending on the intrinsic coordinates $\vect{\xi}_i$ from \cref{eq:hamil:jxi}, which are illustrated in \cref{fig:jncsm:jacobi coordinates}a, with $i=1,\dotsc,A-1$, e.g.
\begin{equation} \label{eq:jncsm:pbasis}
\ket{ (\ldots(\alpha_1,\alpha_2)J_3T_3,\alpha_3)J_4T_4,\ldots,\alpha_{\Am1})JT },
\end{equation}
where $\ket{\alpha_i} \equiv \ket{n_i(l_i s_i)j_i t_i}$ are HO states, associated with coordinates $\vect{\xi}_i$, with $n_i$, $l_i$, $s_i$, and $t_i$ being the radial, orbital, spin, and isospin quantum numbers.
The parentheses in \cref{eq:jncsm:pbasis} indicate the coupling of angular momenta and isospins.
The quantum numbers $J_i$ and $T_i$ ($i=3,\ldots,A$) are angular momentum and isospin quantum numbers of $i$-baryon clusters, with $J_A\equiv J$, $T_A\equiv T$.
The state $\ket{\alpha_1}$ is special because it is a two-body state, where $s_1$ and $t_1$ result from coupling two single-particle (iso\nobreakdash-) spins, while for the other coordinates these quantum numbers are determined by the $(i+1)$\textsuperscript{st} particle.

The basis \eqref{eq:jncsm:pbasis} is truncated by restricting the total number
of HO quanta:
\begin{equation} \label{eq:jncsm:truncation}
\sum_{i=1}^{A-1} (2n_i+l_i) \le \Nmax + N_0
\end{equation}
with $N_0$ the number of HO quanta in the lowest Pauli-allowed state.
\Cref{eq:jncsm:truncation} defines the size of the model space.
Since the NCSM is based on a diagonalization of a matrix representation of the Hamiltonian, calculations are variational and converge to the exact results with increasing $\Nmax$.

\subsection{\label{sec:jncsm:antisymmetrization}Basis antisymmetrization}
The basis states defined in \eqref{eq:jncsm:pbasis} have to be antisymmetrized with respect to the exchanges of all nucleons.
We take the single hyperon as a distinguishable particle and exclude it from the antisymmetrization process.

The antisymmetrization procedure for hypernuclear systems is a straightforward application of the antisymmetrization procedure developed for purely nucleonic systems employing a Jacobi-coordinate HO basis, which is extensively discussed, e.g., in Ref.~\cite{Navratil:1999pw}.
In the following we will only summarize its main points and necessary extensions to hypernuclear systems.

A HO basis fully antisymmetric with respect to exchanges of all $A-1$ nucleons can be obtained via diagonalization of the antisymmetrization operator $\Antisym_{\Am1}$ between the product basis states \eqref{eq:jncsm:pbasis}.
For simplicity, we assume that the hyperon has index $i=A$.
The antisymmetrizer is defined as
\begin{equation} \label{eq:jncsm:A}
\Antisym_{\Am1} = \frac{1}{(A-1)!} \sum_{\pi} \sgn(\pi)\, \Perm_{\pi},
\end{equation}
where the summation extends over all permutations $\pi\in S_{A-1}$ of single-nucleon coordinates, with signature $\sgn(\pi)$ and permutation operator $\Perm_\pi$.
The antisymmetrizer acts as the identity operator on the hyperon coordinate.

The eigenvectors of the antisymmetrizer span two eigen\-spaces, one corresponding to eigenvalue 1 formed by (physical) fully-antisymmetric states and one corresponding to eigenvalue 0 formed by the remaining (unphysical) states.
The antisymmetrizer \eqref{eq:jncsm:A} can be represented as
\begin{equation} \label{eq:jncsm:factA}
\Antisym_{\Am1} = \Antisym_{\Am2} \frac{1}{A-1} \left[ 1-(A-2)\,\Perm_{\Am2,\Am1} \right] \Antisym_{\Am2},
\end{equation}
where the transposition operator $\Perm_{\Am2,\Am1}$ interchanges the coordinates of nucleons $A-2$ and $A-1$.
\Cref{eq:jncsm:factA} provides the basis for an iterative procedure to obtain fully antisymmetrized states from states that are only antisymmetric with respect to exchanges among a subset of the nucleons.

Since the antisymmetrizer \eqref{eq:jncsm:A} is diagonal in all quantum numbers
associated with hyperons, the iterative procedure consists, in the practical
implementation, of first constructing antisymmetrized $(A-1)$-nucleon channels
and then adding a hyperon.
In the simplest $A=3$ hypernuclear case the action of the antisymmetrizer
\eqref{eq:jncsm:A} on the basis states \eqref{eq:jncsm:pbasis} can be easily
evaluated.
% "In the" was repeated
For two nucleons, from the equivalence of the nucleon
interchange and the parity operation in coordinate space follows that
\begin{align}
\Antisym_2 \ket{(\alpha_1,\alpha_2)JT} &
  = \tfrac{1}{2}\left(1-\Perm_{12}\right) \ket{(\alpha_1,\alpha_2)JT}
\notag\\&
  =\tfrac{1}{2}\left[ 1-(-1)^{l_1+s_1+t_1} \right] \ket{(\alpha_1,\alpha_2)JT}. \label{eq:jncsm:A3}
\end{align}
where $\Perm_{12}$ is the transposition operator of nucleons~1 and~2.
Note that the subscripts of the transposition operator refer to nucleons, not Jacobi coordinates and $\Perm_{12}$ acts on the first Jacobi coordinate, with corresponding state $\ket{\alpha_1}$, only.
From here it follows that the antisymmetry with respect to exchange of the nucleons in the state $\ket{\alpha_1}$ is simply achieved by restricting the relative quantum numbers of the two-nucleon HO channels by $(-1)^{l_1+s_1+t_1}=-1$.
The result of \cref{eq:jncsm:A3} together with the representation of the antisymmetrizer in \cref{eq:jncsm:factA} is used to initiate the iterative procedure of constructing the antisymmetrized basis for a larger number of particles, by adding one nucleon at a time. Details of the antisymmetrization procedure, together with matrix elements of the antisymmetrizer for arbitrary number of nucleons, can be found in Ref.~\cite{Navratil:1999pw}. The resulting states, labeled as
\begin{equation} \label{eq:jncsm:abasis}
\ket{ (N_{\Am1}i_{\Am1}J_{\Am1}T_{\Am1},\alpha_{\Am1})JT},
\end{equation}
are antisymmetric with respect to exchanges of all nucleons. In \eqref{eq:jncsm:abasis} we adopt the notation of Ref.~\cite{Navratil:1999pw} for the antisymmetric states of $A-1$ nucleons coupled to total angular momentum $J_{\Am1}$ and isospin $T_{\Am1}$ as
\begin{equation} \label{eq:jncsm:anstate}
\ket{N_{\Am1} i_{\Am1} J_{\Am1} T_{\Am1}},
\end{equation}
where the quantum number $i_{\Am1}$ distinguishes between different antisymmetric states with the same quantum numbers $N_{\Am1},J_{\Am1},T_{\Am1}$.
The quantum number $N_{\Am1}$, specified below, is the total number of HO quanta in the state.
Following Eq.~\eqref{eq:jncsm:factA} the state \eqref{eq:jncsm:anstate} can be expanded in states containing antisymmetric subcluster of $A-2$ nucleons and one nucleon as
\begin{align}
\MoveEqLeft\ket{N_{\Am1} i_{\Am1} J_{\Am1} T_{\Am1}}=\sum_{N_{\Am2} i_{\Am2} J_{\Am2} T_{\Am2}}\sum_{\alpha_{\Am2}}
\notag\\&
  \braket{(N_{\Am2} i_{\Am2} J_{\Am2} T_{\Am2},\alpha_{\Am2})J_{\Am1}T_{\Am1}| N_{\Am1} i_{\Am1} J_{\Am1} T_{\Am1}}
\notag\\&
  \times\ket{(N_{\Am2} i_{\Am2} J_{\Am2} T_{\Am2},\alpha_{\Am2})J_{\Am1}T_{\Am1}}, \label{eq:jncsm:anstateexp}
\end{align}
where the expansion coefficients obtained from the eigenvectors of the
antisymmetrizer are the coefficients of fractional parentage and
$N_{\Am1} = N_{\Am2}+2n_{\Am2}+l_{\Am2}$ (in the two-nucleon case $N_2 = 2n_1+l_1$).

\subsection{Evaluation of interaction matrix elements} \label{sec:jncsm:interaction}
\begin{figure}
  \centering
  \begin{tikzpicture}[x=1.5cm]
    \node[draw, circle] (core) at (0,0) {$A-3$};
    \node[draw, circle] (n2) at ($ (core) + (-30:1.3) $) {N};
    \node[draw, circle] (n1) at ($ (core) + (10:1.3) $) {N};
    \node[draw, circle] (hyp) at ($ (core) + (50:1.7) $) {Y};
    \draw (core) -- (n2) node[pos=0.7, auto=right, inner sep=2pt] {$\xi_{\Am3}$};
    \draw ($ (core)!.5!(n2) $) -- (n1) node[midway, auto=left, inner sep=2pt] {$\xi_{\Am2}$};
    \draw ($ (core)!.5!(n2)!.3!(n1) $) -- (hyp) node[midway, auto=left, inner sep=2pt] {$\xi_{\Am1}$};
    \node[anchor=north] at (current bounding box.south) {(a)};
  \end{tikzpicture}\hspace{2em}
  \begin{tikzpicture}[x=1.5cm]
    \node[draw, circle] (core) at (0,0) {$A-2$};
    \node[draw, circle] (n1) at ($ (core) + (-30:1.3) $) {N};
    \node[draw, circle] (hyp) at ($ (core) + (50:1.7) $) {Y};
    \draw (core) -- ($ (hyp)!.35!(n1) $) node[near end, auto=left, inner sep=2pt] {$\eta_{\Am2}$};
    \draw (n1) -- (hyp) node[midway, auto=right, inner sep=2pt] {$\eta_{\Am1}$};
    \node[anchor=north] at (current bounding box.south) {(b)};
  \end{tikzpicture}\\
  \begin{tikzpicture}[x=1.5cm]
    \node[draw, circle] (core) at (0,0) {$A-3$};
    \node[draw, circle] (n1) at ($ (core) + (-30:1.3) $) {N};
    \node[draw, circle] (n2) at ($ (core) + (10:1.3) $) {N};
    \node[draw, circle] (hyp) at ($ (core) + (50:1.7) $) {Y};
    \draw (core) -- ($ (n1)!.5!(n2) $) node[pos=0.7, auto=right, inner sep=2pt] {$\rho_{\Am3}$};
    \draw (n1) -- (n2) node[pos=1, auto=right, inner sep=2pt] {$\rho_{\Am2}$};
    \draw ($ (n1)!.5!(n2)!.5!(core) $) -- (hyp) node[midway, auto=left, inner sep=2pt] {$\xi_{\Am1}$};
    \node[anchor=north] at (current bounding box.south) {(c)};
  \end{tikzpicture}\hspace{2em}
  \begin{tikzpicture}[x=1.5cm]
    \node[draw, circle] (core) at (0,0) {$A-4$};
    \node[draw, circle] (n1) at ($ (core) + (-30:1.6) $) {N};
    \node[draw, circle] (n2) at ($ (core) + (10:1.3) $) {N};
    \node[draw, circle] (n3) at ($ (core) + (-80:1) $) {N};
    \node[draw, circle] (hyp) at ($ (core) + (50:1.7) $) {Y};
    \draw (core) -- ($ (n1)!.5!(n2)!.3!(n3) $) node[pos=0.4, auto=left, inner sep=2pt] {$\zeta_{\Am4}$};
    \draw (n3) -- ($ (n1)!.5!(n2) $) node[pos=0.1, auto=right, inner sep=2pt] {$\zeta_{\Am3}$};
    \draw (n1) -- (n2) node[pos=0.3, auto=right, inner sep=2pt] {$\rho_{\Am2}$};
    \draw ($ (n1)!.5!(n2)!.3!(n3)!.6!(core) $) -- (hyp) node[pos=0.7, auto=left, inner sep=2pt] {$\xi_{\Am1}$};
    \node[anchor=north] at (current bounding box.south) {(d)};
  \end{tikzpicture}
  \caption{\label{fig:jncsm:jacobi coordinates}Illustration of the Jacobi coordinates used for (a) antisymmetrization and embedding of the (b) \YN{}, (c) \NN{}, and (d) \NNN{} interactions.}
\end{figure}
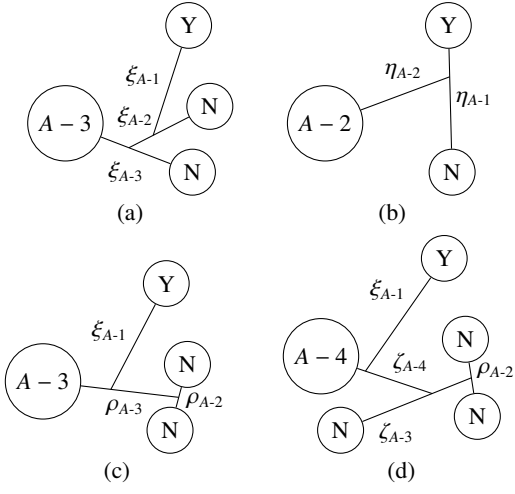
%
% embedding of V_NY
Since we treat the hyperon as a distinguishable particle the set of Jacobi coordinates \eqref{eq:hamil:jxi} and the associated antisymmetric basis constructed in the previous section are not convenient for the evaluation of two- and three-body interaction matrix elements.
A different set of Jacobi coordinates suitable when matrix elements of the hyperon-nucleon interaction need to be evaluated is obtained by keeping the coordinates $\vect{\xi}_0,\dotsc, \vect{\xi}_{A-3}$ from the set
\eqref{eq:hamil:jxi} and introducing two new coordinates (cf.\ \cref{fig:jncsm:jacobi coordinates}b):
\begin{subequations}\label{eq:jncsm:jeta}
\begin{align}
\vect{\eta}_{\Am2} &= \sqrt{\frac{\submass{1}{\Am2}\submass{\Am1}{A}}{\submass{1}{A}}}\left(
\frac{1}{\submass{1}{\Am2}}\sum_{i=1}^{A-2}\sqrt{m_i}\vect{x}_i\right. \notag\\*
&\pheq - \left.\frac{1}{\submass{\Am1}{A}}\sum_{i=A-1}^{A} \sqrt{m_i}\vect{x}_i \right),\\
\vect{\eta}_{\Am1} &=
\sqrt{\frac{m_{\Am1}m_{A}}{\submass{\Am1}{A}}}\left(\frac{1}{\sqrt{m_{\Am1}}}\vect{x}_{\Am1}
- \frac{1}{\sqrt{m_A}} \vect{x}_A \right).
\end{align}
\end{subequations}
Thus, the antisymmetrized states \eqref{eq:jncsm:abasis} need to be expanded in a HO basis consisting of antisymmetrized (for $A>3$) states of $A-2$ nucleons and nucleon-hyperon states:
\begin{align} % for evaluation of V_NY
\label{eq:jncsm:reta}
\MoveEqLeft\ket{ ( N_{\Am1}i_{\Am1}J_{\Am1}T_{\Am1},\alpha_{\Am1}) JT} = \sum_{N_{\Am2}i_{\Am2}J_{\Am2}T_{\Am2}} \sum_{\alpha_{\Am2}} \sum_{nlsjt\pcl} \sum_{\Ncal\Lcal\Jcal L}
\kronecker{\pcl}{t_{\Am1}} \notag\\*
&\braket{ (N_{\Am2}i_{\Am2}J_{\Am2}T_{\Am2},\alpha_{\Am2})J_{\Am1}T_{\Am1} | N_{\Am1}i_{\Am1}J_{\Am1}T_{\Am1} } \notag\\
&\times \hat{T}_{\Am1} \hat{t} (-1)^{T_{\Am2}+\frac{1}{2}+t_{\Am1}+T}
\sixj{T_{\Am2}}{\frac{1}{2}}{T_{\Am1}}{t_{\Am1}}{T}{t}
\notag\\
&\times \jhat_{\Am1} \hat{J}_{\Am1} \hat{L}^2 \jhat_{\Am2} \hat{s}
\hat{\Jcal}\jhat
(-1)^{J_{\Am2}+j_{\Am2}+J_{\Am1}+J+\Lcal+j+l_{\Am2}+l_{\Am1}+s}
\notag\\
&\times
\ninej{l_{\Am2}}{\tfrac{1}{2}}{j_{\Am2}}{l_{\Am1}}{\tfrac{1}{2}}{j_{\Am1}}{L}{s}{\Jcal}
\sixj{J_{\Am2}}{j_{\Am2}}{J_{\Am1}}{j_{\Am1}}{J}{\Jcal}
\sixj{\Lcal}{l}{L}{s}{\Jcal}{j}
\notag\\
&\times (-1)^{\Lcal + l - L} \hob{\Ncal\Lcal}{nl}{n_{\Am1}l_{\Am1}}{n_{\Am2}l_{\Am2}}{L}{\frac{(\Am2)m_{\pcl}}{(\Am1)m_N+m_{\pcl}}}
\notag\\
&\times\ket{ (
N_{\Am2}i_{\Am2}J_{\Am2}T_{\Am2},(nlsjt\pcl,\Ncal\Lcal)\Jcal) JT },
\end{align}
where the objects in curly braces are Wigner $6j$ and $9j$ symbols. The coefficients $\hob{n_1l_1}{n_2l_2}{NL}{nl}{\lambda}{d}$ are harmonic-oscillator brackets (HOBs) that mediate the transformation between two coordinate sets $\vect{x}_1,\vect{x}_2$ and $\vect{X},\vect{x}$, e.g., the coordinates $\vect{x}_1,\vect{x}_2$ and $\vect{\xi}_0,\vect{\xi}_1$ from \cref{eq:hamil:2b jacobi}.
It gives the overlap between HO states, coupled to total orbital angular momentum $\lambda$, defined with respect to these coordinate sets \cite{Talmi1952,Moshinsky1959}. The formulation of the HOBs that we employ (cf.\ Ref.~\cite{Kamuntavicius2001}) requires the two sets to be connected by an orthogonal transformation
\begin{equation}\label{eq:jncsm:orthogonal transformation}
  \begin{pmatrix}
    \vect{X} \\ \vect{x}
  \end{pmatrix}
  =
  \begin{pmatrix}
    \sqrt{\frac{d}{1+d}} & \sqrt{\frac{1}{1+d}} \\
    \sqrt{\frac{1}{1+d}} & -\sqrt{\frac{d}{1+d}}
  \end{pmatrix}
  \begin{pmatrix}
    \vect{x}_1 \\ \vect{x}_2
  \end{pmatrix}
\end{equation}
with the parameter $d$ defining the transformation.%
\footnote{The HO wavefunction used in Ref.~\cite{Kamuntavicius2001} carries an additional factor of $(-1)^n$, so that our HOBs differ by a phase $(-1)^{n+N+n_1+n_2}$. Also, Ref.~\cite{Navratil:1999pw} uses the definition of \cite{Trlifaj1972}, which differs by a phase $(-1)^{L+l-\lambda}$ and $d\mapsto 1/d$.}
In Eq.~\eqref{eq:jncsm:reta}, an orthogonal transformation between coordinates $\vect{\xi}_{\Am2}$, $\vect{\xi}_{\Am2}$ and $\vect{\eta}_{\Am2}$, $\vect{\eta}_{\Am1}$ was employed.
In the state
\begin{equation} \label{eq:jncsm:beta}
\ket{ ( N_{\Am2}i_{\Am2}J_{\Am2}T_{\Am2},(nlsjt\pcl,\Ncal\Lcal)\Jcal) JT }
\end{equation}
the antisymmetrized nucleon subcluster $\ket{N_{\Am2}i_{\Am2}J_{\Am2}T_{\Am2}}$ depends on the Jacobi coordinates $\vect{\xi}_{1},\dotsc,\vect{\xi}_{\Am3}$.
In the relative hyperon--nucleon HO states $\ket{nlsjt\pcl}$ depending on the
coordinate $\vect{\eta}_{\Am1}$ the additional quantum number $\pcl$
distinguishes channels containing $\pcl=\Lambda$ and $\pcl=\Sigma$
hyperons with mass $m_{\pcl}$. The HO state $\ket{\Ncal\Lcal}$ depends on the Jacobi
coordinate $\vect{\eta}_{\Am2}$ and describes the relative motion of the two
subclusters. With the help of expansion \eqref{eq:jncsm:reta} it is
straightforward to evaluate the \NY{} interaction matrix elements in the
antisymmetrized basis \eqref{eq:jncsm:abasis},
\begin{equation} \label{eq:jncsm:v2ny}
\braket{\op{V}^{[2]}_{\NY}} =
(A-1)\braket{\op{V}^{[2]}_{\NY}\left(\vect{\eta}_{\Am1}\right)},
\end{equation}
where the matrix element on the right-hand side is diagonal in all quantum numbers of the state \eqref{eq:jncsm:beta} except in $n,l,\pcl$ for isospin-conserving interactions and the interaction operator acts only on the state given by the coordinate in parentheses.

% embedding of V_NN
Similarly, when matrix elements of two-body nucleon-nucleon interactions are to
be evaluated it is convenient to define a new set of Jacobi coordinates by keeping
$\vect{\xi}_0, \dotsc$, $\vect{\xi}_{\Am4}$, $\vect{\xi}_{\Am1}$ from the set
\eqref{eq:hamil:jxi} and introducing two new coordinates (cf.\ \cref{fig:jncsm:jacobi coordinates}c)
\begin{subequations} \label{eq:jncsm:jrho} % for evaluation of V_NN
\begin{align}
\vect{\rho}_{\Am3} &=\sqrt{\frac{\submass{1}{\Am3}\submass{\Am2}{\Am1}}{\submass{1}{\Am1}}} \left(
\frac{1}{\submass{1}{\Am3}}\sum_{i=1}^{A-3}\sqrt{m_i}\vect{x}_i \right. \notag\\*
&\pheq- \left. \frac{1}{\submass{\Am2}{\Am1}}\sum_{i=A-2}^{A-1}\sqrt{m_i}\vect{x}_i \right), \\
\vect{\rho}_{\Am2} &=\sqrt{\frac{m_{\Am2}m_{\Am1}}{\submass{\Am2}{\Am1}}} \left(
\frac{1}{\sqrt{m_{\Am2}}} \vect{x}_{\Am2} - \frac{1}{\sqrt{m_{\Am1}}} \vect{x}_{\Am1} \right),
\end{align}
\end{subequations}
and expanding the antisymmetrized states \eqref{eq:jncsm:abasis} in a basis
containing HO states of relative two-nucleon channels,
\begin{align} % for evaluation of V_NN
\MoveEqLeft\ket{ (N_{\Am1} i_{\Am1} J_{\Am1} T_{\Am1}, \alpha_{\Am1}) JT } = \sum_{\substack{N_{\Am2}i_{\Am2}J_{\Am2}T_{\Am2} \\ N_{\Am3}i_{\Am3}J_{\Am3}T_{\Am3}}} \sum_{\substack{\alpha_{\Am2} \\ \alpha_{\Am3}}} \sum_{\Ncal\Lcal\Jcal L} \sum_{nlsjt}
\notag\\&
\braket{(N_{\Am2}i_{\Am2}J_{\Am2}T_{\Am2},\alpha_{\Am2})J_{\Am1}T_{\Am1}\vert N_{\Am1} i_{\Am1} J_{\Am1} T_{\Am1}}
\notag\\&\times
\braket{(N_{\Am3}i_{\Am3}J_{\Am3}T_{\Am3},\alpha_{\Am3})J_{\Am2}T_{\Am2} \vert N_{\Am2}i_{\Am2}J_{\Am2}T_{\Am2}}
\notag\\&\times
\jhat_{\Am2}\jhat_{\Am3}\hat{\Jcal}\hat{J}_{\Am2} \jhat\,\hat{s}\, (-1)^{j_{\Am3}+j_{\Am2}+J_{\Am3}+J_{\Am1}+j+\Lcal+s+l_{\Am2}+l_{\Am3}}
\notag\\&\times
(-1)^{T_{\Am3}+T_{\Am1}+1}
\sixj{T_{\Am3}}{\tfrac{1}{2}}{T_{\Am2}}{\tfrac{1}{2}}{T_{\Am1}}{t}
\ninej{l_{\Am3}}{\tfrac{1}{2}}{j_{\Am3}}{l_{\Am2}}{\tfrac{1}{2}}{j_{\Am2}}{L}{s}{\Jcal}
\notag\\&\times
\sixj{J_{\Am3}}{j_{\Am3}}{J_{\Am2}}{j_{\Am2}}{J_{\Am1}}{\Jcal}
\sixj{s}{l}{j}{\Lcal}{\Jcal}{L}
\notag\\&\times
\hat{L}^2 (-1)^{\Lcal+l-L}
\hob{\Ncal\Lcal}{nl}{n_{\Am2}l_{\Am2}}{n_{\Am3}l_{\Am3}}{L}{\frac{A-3}{A-1}}
\notag\\&\times
\ket{ ((N_{\Am3}i_{\Am3}J_{\Am3}T_{\Am3},(nlsjt,
\Ncal\Lcal)\Jcal) J_{\Am1}T_{\Am1},\alpha_{\Am1})JT }. \label{eq:jncsm:rrho}
\end{align}
In Eq.~\eqref{eq:jncsm:rrho} we used an orthogonal transformation between the
coordinates $\vect{\xi}_{\Am3},\vect{\xi}_{\Am2}$ and
$\vect{\rho}_{\Am3},\vect{\rho}_{\Am2}$ and the states
\begin{equation} \label{eq:jncsm:brho}
\ket{ ((N_{\Am3}i_{\Am3}J_{\Am3}T_{\Am3},(nlsjt, \mathcal{NL})\Jcal) J_{\Am1}T_{\Am1},\alpha_{\Am1})JT }
\end{equation}
contain antisymmetric $(A-3)$-nucleon subcluster HO states
$\ket{N_{\Am3}i_{\Am3}J_{\Am3}T_{\Am3}}$ depending on Jacobi coordinates
$\vect{\xi}_1, \ldots$, $\vect{\xi}_{\Am4}$ and relative two-nucleon HO states
$\ket{nlsjt}$ depending on the coordinate $\vect{\rho}_{\Am2}$.
The HO states $\ket{\mathcal{NL}}$ depending on coordinate
$\vect{\rho}_{\Am3}$ describe the relative motion of the two-nucleon and the
$(A-4)$-nucleon clusters and the states
$\ket{\alpha_{\Am1}}$ depending on $\vect{\xi}_{\Am1}$ correspond to the relative
motion of the hyperon with respect to the c.m.\ of the two subclusters.
By using the expansion \eqref{eq:jncsm:rrho} we can evaluate the matrix
elements of the \NN{} interaction in the antisymmetrized basis
\eqref{eq:jncsm:abasis} simply as
\begin{equation} \label{eq:jncsm:v2nn}
\braket{\op{V}^{[2]}_{\NN}} =
\frac{1}{2}(A-1)(A-2)\braket{\op{V}^{[2]}_{\NN}\left(\vect{\rho}_{\Am2}\right)},
\end{equation}
where the matrix element on the right hand side is diagonal in all quantum
numbers of the states \eqref{eq:jncsm:brho} except in $n,l$ for
isospin-conserving interactions.

% embedding of V_NNN
Finally, for calculations of $A>6$ hypernuclear systems with \NNN{} interactions
it is convenient to separate a three-nucleon subcluster and use the Jacobi coordinates
$\vect{\xi}_0, \ldots, \vect{\xi}_{\Am5}, \vect{\rho}_{\Am2},\vect{\xi}_{\Am1}$ from
the previous sets together with two new coordinates (cf.\ \cref{fig:jncsm:jacobi coordinates}d)
\begin{subequations} \label{eq:jncsm:jzeta} % for evaluation of V_NNN
\begin{align}
\vect{\zeta}_{\Am4} &= \sqrt{\frac{\submass{1}{\Am4}\submass{\Am3}{\Am1}}{\submass{1}{\Am1}}}\left( \frac{1}{\submass{1}{\Am4}}\sum_{i=1}^{\Am4}\sqrt{m_i}\vect{x}_i \right. \notag\\
&\pheq- \left. \frac{1}{\submass{\Am3}{\Am1}} \sum_{i=A-3}^{A-1}\sqrt{m_i}\vect{x}_i \right), \\
\vect{\zeta}_{\Am3} &= \sqrt{\frac{m_{\Am3}\submass{\Am2}{\Am1}}{\submass{\Am3}{\Am1}}}
\left( \frac{1}{\sqrt{m_{\Am3}}}\vect{x}_{\Am3}
-\frac{1}{\submass{\Am2}{\Am1}}\sum_{i=A-2}^{A-1}\sqrt{m_i}\vect{x}_i \right).
\end{align}
\end{subequations}
Using an orthogonal transformation between the coordinates
$\vect{\xi}_{\Am4},\vect{\rho}_{\Am3}$ and
$\vect{\zeta}_{\Am4},\vect{\zeta}_{\Am3}$, the basis
states \eqref{eq:jncsm:abasis} can be expanded in the following way:
\begin{widetext}
\begin{align} % for evaluation of V_NNN
\MoveEqLeft\ket{ (N_{\Am1} i_{\Am1} J_{\Am1} T_{\Am1},\alpha_{\Am1}) JT }  =
\sum_{\substack{N_{\Am3}i_{\Am3}J_{\Am3}T_{\Am3} \\  N_{\Am4} i_{\Am4} J_{\Am4} T_{\Am4}}}
\sum_{\substack{n_{\Am4} l_{\Am4} j_{\Am4} \\ n_{\Am4}' l_{\Am4}' j_{\Am4}'}}
\sum_{\Ncal\Lcal\Jcal KL}
\sum_{nlsjt}\sum_{N_3i_3J_3T_3}
\notag\\&
\braket{
((N_{\Am3}i_{\Am3}J_{\Am3}T_{\Am3}, (nlsjt,\Ncal\Lcal)\Jcal)J_{\Am1}T_{\Am1},\alpha_{\Am1})JT \vert
(N_{\Am1}i_{\Am1}J_{\Am1}T_{\Am1},\alpha_{\Am1})JT}
\notag\\&\times
\braket{ ( N_{\Am4} i_{\Am4} J_{\Am4} T_{\Am4}, n_{\Am4} l_{\Am4} j_{\Am4})J_{\Am3}T_{\Am3}
\vert N_{\Am3} i_{\Am3} J_{\Am3} T_{\Am3}}
\braket{(nlsjt,n_{\Am4}' l_{\Am4}' j_{\Am4}')J_3T_3 \vert N_3i_3J_3T_3}
\jhat_{\Am4}\jhat_{\Am4}'\hat{\Jcal}\hat{\Jcal}'
\hat{J}_{\Am3}\hat{J}_3
\notag\\&\times
(-1)^{\Jcal'+J_{\Am4}+J_{\Am1}+j+l_{\Am4}+\tfrac{1}{2}}
(-1)^{T_{\Am4}+T_3+T_{\Am1}}\hat{T}_{\Am3}\hat{T}_3
\sixj{T_{\Am4}}{\tfrac{1}{2}}{T_{\Am3}}{t}{T_{\Am1}}{T_3}
\sixj{J_{\Am4}}{J_{\Am1}}{\Jcal}{\Jcal'}{j_{\Am4}}{J_{\Am3}}
\hat{K}^2(-1)^K
\ninej{j_{\Am4}'}{J_3}{j}{l_{\Am4}'}{\Lcal}{L}{\tfrac{1}{2}}{\Jcal}{K}
\notag\\&\times
\sixj{\Jcal'}{l_{\Am4}}{K}{\tfrac{1}{2}}{\Jcal}{j_{\Am4}}
\sixj{\Jcal'}{l_{\Am4}}{K}{L}{j}{\Lcal'}
\hat{L}^2 (-1)^{\Lcal + l_{\Am4}'}
\hob{\Ncal\Lcal}{n_{\Am4}'l_{\Am4}'}{n_{\Am4}l_{\Am4}}{\Ncal'\Lcal'}{L}{\frac{A-1}{2(A-4)}}
\notag\\&\times
\ket{((N_{\Am4}i_{\Am4}J_{\Am4}T_{\Am4},(N_{3}i_{3}J_{3}T_{3},\Ncal\Lcal)\Jcal)
J_{\Am1}T_{\Am1},\alpha_{\Am1})JT }, \label{eq:jncsm:rzeta}
\end{align}
\end{widetext}
where the basis expansion coefficient obtained from Eq.~\eqref{eq:jncsm:rrho}
was used. In the state
\begin{equation} \label{eq:jncsm:bzeta}
\ket{((N_{\Am4}i_{\Am4}J_{\Am4}T_{\Am4},(N_{3}i_{3}J_{3}T_{3},\mathcal{NL})\Jcal)
J_{\Am1}T_{\Am1},\alpha_{\Am1})JT},
\end{equation}
the HO state $\ket{N_{\Am4}i_{\Am4}J_{\Am4}T_{\Am4}}$ corresponding to the
antisymmetric $(A-4)$-nucleon cluster depends on the Jacobi coordinates
$\vect{\xi}_{1},\ldots,\vect{\xi}_{\Am5}$ and the three-nucleon HO channels
$\ket{N_{3}i_{3}J_{3}T_{3}}$ depend on the coordinates $\vect{\zeta}_{\Am3}$ and
$\vect{\rho}_{\Am2}$. The HO state $\ket{\Ncal\Lcal}$ associated with coordinate
$\vect{\zeta}_{\Am4}$ describes the relative motion of the $(A-4)$- and
3-nucleon clusters and the HO state $\ket{\alpha_{\Am1}}$ depending on
coordinate $\vect{\xi}_{\Am1}$ corresponds to the motion of hyperon relative to
the c.m.\ of the two clusters.
The matrix elements of \NNN{} interactions between the states
\eqref{eq:jncsm:abasis} are easily evaluated using the expansion
\eqref{eq:jncsm:rzeta} as
\begin{equation} \label{eq:jncsm:v3nnn}
\braket{\op{V}^{[3]}_{\NNN}} = \frac{1}{6} (A-1)(A-2)(A-3)
\braket{\op{V}^{[3]}_{\NNN}(\vect{\zeta}_{\Am3},\vect{\rho}_{\Am2})},
\end{equation}
where the matrix element on the right hand side is diagonal in all quantum
numbers of the state \eqref{eq:jncsm:bzeta} except for $N_3$ and $i_3$, for
isospin-invariant interactions.

\section{\label{sec:itncsm}Single-Particle-Coordinate Formulation}
With increasing particle number the efficiency of the Ja\-cobi-coor\-dinate formulation is hampered by the increasing computational effort for the antisymmetrization of basis states and the evaluation of interaction matrix elements.
A formulation in terms of single-particle coordinates is then more efficient because one can use a trivially-antisymmetric Slater determinant basis in which the calculation of many-body matrix elements is simple.
The tradeoff is that the inclusion of the center-of-mass degrees of freedom increases the basis dimension dramatically and that the Slater determinant basis does not fully exploit the other symmetries of the Hamiltonian such as rotational invariance.
However, when using the $\Nmax$ truncation a decoupling between c.m.\ and intrinsic degrees of freedom is retained.
An additional importance-truncation scheme allows for an efficient treatment of the large model spaces.
The following sections discuss the transformation of interaction matrix elements to single-particle coordinates, the NCSM model space and the importance-truncation scheme.

\subsection{\label{sec:itncsm:tm transformation}Transformation to Single-Particle Coordinates}
For the single-particle-coordinate formulation, we have to transform the interaction matrix elements from the two-body Jacobi basis into a HO basis with quantum numbers given with respect to single-particle coordinates.
The transformation of the \NNN{} matrix elements is presented in Ref.~\cite{Roth2014}, in the following we show the transformation of the two-body interaction for particles with unequal masses.
We derive the full transformation, generalizing the result from Refs.~\cite{Barrett1971,Roth2006}.
The transformation between the two-body Jacobi and single-particle coordinates itself is achieved by a HOB.
The parameter defining the transformation can be read off from \cref{eq:hamil:2b jacobi} as $d=m_1/m_2$.

First, we express a $J$\nobreakdash-coupled state $\keta{(\Jred{a}\Jred{b})JM_J}$ in terms of relative HO states.
The shorthands $\Jred{a}=\{n_a(l_as_a)j_a\pcl_a\}$, etc., collect the single-particle quantum numbers.
Working in particle basis, the antisymmetrization of this state is more complicated than that of the $JT$\nobreakdash-coupled states described in~\cite{Barrett1971,Roth2006} because antisymmetrization cannot be achieved via a selection rule on angular-momentum and isospin quantum numbers alone.
Similar to the antisymmetrization of the relative states we have
\begin{align}
 \Antisym\ket{(\Jred{a}\Jred{b})JM_J} &= \tfrac12 \bigl(\ket{(\Jred{a}\Jred{b})JM_J} - (-1)^{j_a+j_b-J}\ket{(\Jred{b}\Jred{a})JM_J}\bigr)
\shortintertext{and}
 \keta{(\Jred{a}\Jred{b})JM_J} &= \sqrt2 \mathcal{N} \Antisym\ket{(\Jred{a}\Jred{b})JM_J}
    \label{eq:itncsm:antisymmetrized ket}
\end{align}
with $\mathcal{N} = (1 + \kronecker{\Jred{a}}{\Jred{b}})^{-1/2}$ a normalization coefficient.

Using the relations between different angular-momentum coupling schemes from \cite{Varshalovich1988}, the defining relations of the HOBs \cite{Kamuntavicius2001}, and assuming that all particles have spin $1/2$, we are able to express a $J$\nobreakdash-coupled state in terms of relative and center-of-mass HO states:
\begin{align}
 \MoveEqLeft\keta{(\Jred{a}\Jred{b})JM_J} =
  \mathcal{N}(1+\kronecker{\pcl_a}{\pcl_b})^{\frac12}
  \sum_j \sum_{\substack{N\Lcal \\ n\lambda}} \sum_{LS} \sum_{m_\Lcal m_j}
  \hat{L}^2\hat{S}\jhat\jhat_a\jhat_b
  \notag\\&\times
  (-1)^{\Lcal+\lambda+S+J}
    \ninej{l_a}{s_a}{j_a}{l_b}{s_b}{j_b}{L}{S}{J}
    \sixj{\Lcal}{\lambda}{L}{S}{J}{j}
    \clebsch{\Lcal}{m_\Lcal}{j}{m_j}{J}{M_J}
  \notag\\&\times
  \hob{N\Lcal}{n\lambda}{n_1l_1}{n_2l_2}{L}{d}
  \notag\\&\times
  \ket{N\Lcal m_\Lcal}\keta{[n\lambda,(s_as_b)S]jm_j,\pcl_a\pcl_b},
  \label{eq:itncsm:ket antisymmetrized transformation}
\end{align}
with Kronecker delta $\kronecker{\pcl_a}{\pcl_b}$.
The symbol $\clebsch{j_a}{m_a}{j_b}{m_b}{J}{M_J}$ denotes a Clebsch-Gordan coefficient; $N,\Lcal,m_\Lcal$ are the center-of-mass radial and orbital quantum numbers.

The same transformation \eqref{eq:itncsm:ket antisymmetrized transformation} can be applied to the bra state and as our final result we get the expression for a matrix element
\begin{align}
 \MoveEqLeft\abraketa{(\Jred{a}'\Jred{b}')JM_J|\Valpha|(\Jred{a}\Jred{b})JM_J} = \notag\\
 &\mathcal{N}\mathcal{N}' (1+\kronecker{\pcl_a}{\pcl_b})^{\frac12} (1+\kronecker{\pcl'_a}{\pcl'_b})^{\frac12}
  \sum_{j}
  \sum_{N\Lcal}
  \sum_{\substack{n\lambda \\ n'\lambda'}}
  \sum_{\substack{LS \\ L'S'}}
  \notag\\
 &\times
  \hat{L}^2\hat{L}'^2\hat{S}\hat{S}'\jhat^2\jhat_a\jhat'_a\jhat_b\jhat'_b
  (-1)^{S+S'}
  \notag\\
 &\times
  \ninej{l_a}{s_a}{j_a}{l_b}{s_b}{j_b}{L}{S}{J}
  \ninej{l'_a}{s'_a}{j'_a}{l'_b}{s'_b}{j'_b}{L'}{S'}{J}
  \sixj{\Lcal}{\lambda}{L}{S}{J}{j}
  \sixj{\Lcal}{\lambda'}{L'}{S'}{J}{j}
  \notag\\
 &\times
  \hob{N\Lcal}{n\lambda}{n_al_a}{n_bl_b}{L}{d}
  \notag\\
 &\times
  \hob{N\Lcal}{n'\lambda'}{n'_al'_a}{n'_bl'_b}{L'}{d'}
  \notag\\
 &\times
  \abraketa{[n'\lambda',(s'_as'_b)S']j,\pcl'_a\pcl'_b|\Valpha|[n\lambda,(s_as_b)S]j,\pcl_a\pcl_b},
  \label{eq:itncsm:antisymmetric matrix element transformation}
\end{align}
where $d'$ refers to the mass ratio of the particles $\pcl_a',\pcl_b'$ and we used that $\Valpha$ only acts on the relative state.
Note that this expression can also be used to transform other rotationally-invariant operators that act only on relative coordinates.
Expressions for (reduced) matrix elements of spherical tensors of nonzero rank can be derived from \cref{eq:itncsm:ket antisymmetrized transformation} in a straight-forward manner.

We can recover $m$\nobreakdash-scheme matrix elements from these matrix elements by removing the $J$\nobreakdash-coupling:
\begin{align}
 \MoveEqLeft\abraketa{\Jred{a}'m'_a,\Jred{b}'m'_b|\Valpha|\Jred{a}m_a,\Jred{b}m_b} \notag\\*
 &= \sum_{JM_J}\clebsch{j'_a}{m'_a}{j'_b}{m'_b}{J}{M_J}\clebsch{j_a}{m_a}{j_b}{m_b}{J}{M_J} \notag\\
 &\pheq{}\times\mathcal{N}^{-1}\mathcal{N}'^{-1}\abraketa{(\Jred{a}'\Jred{b}')JM_J|\Valpha|(\Jred{a}\Jred{b})JM_J}.
\end{align}
These two-body $m$\nobreakdash-scheme matrix elements enter into the calculation of many-body matrix elements between Slater determinants.
The normalization factors $\mathcal{N}$ are often omitted in actual calculations because they appear as a prefactor in the transformation \eqref{eq:itncsm:antisymmetric matrix element transformation} and their inverse multiplies the coupled matrix element during decoupling so that they have no net effect.

\subsection{\label{sec:itncsm:modelspace}NCSM Model Space}
The NCSM model space $\modelspace$ is spanned by a basis of Slater determinants of HO states,
\begin{equation}
 \ket{\phi_i} = \keta{n_1(l_1s_1)j_1m_1\pcl_1,\dotsc,n_A(l_As_A)j_Am_A\pcl_A}.
\end{equation}
This basis is called an $m$\nobreakdash-scheme basis due to the explicit an\-gular-momen\-tum projection quantum numbers.
Since the Hamiltonian of the nuclear system is rotationally invariant, the nuclear energy levels are degenerate with respect to the total angular momentum projection
\begin{equation}
 M_J=\sum_{i=1}^A m_i
\end{equation}
and we can restrict the space to a specific $M_J$---typically the smallest allowed value $M_J=0$ ($M_J=\pm1/2$) for even (odd) number of particles.
Exploiting the parity symmetry of the Hamiltonian, we also restrict the model space to basis states of either odd or even parity.

We additionally truncate the model space by introducing a parameter $\Nmax$ with
\begin{equation} \label{eq:itncsm:nmax truncation}
  \sum_{i=1}^{A} (2n_i + l_i) \le \Nmax + N_0,
\end{equation}
where $N_0$ is the total number of excitation quanta in the determinant constructed from the lowest Pauli-allowed single-particle states.

The hyperon--nucleon interaction allows changing the identity of particles via $\Lambda N$--$\Sigma N$ coupling.
Hence, to get the full model space we generate Slater determinants with all combinations of particles allowed by the total strangeness $\Strng$ and charge $Q$ (or, equivalently, total isospin projection $M_T$) of the system under consideration.

\begin{figure}
  \centering
  \includegraphics{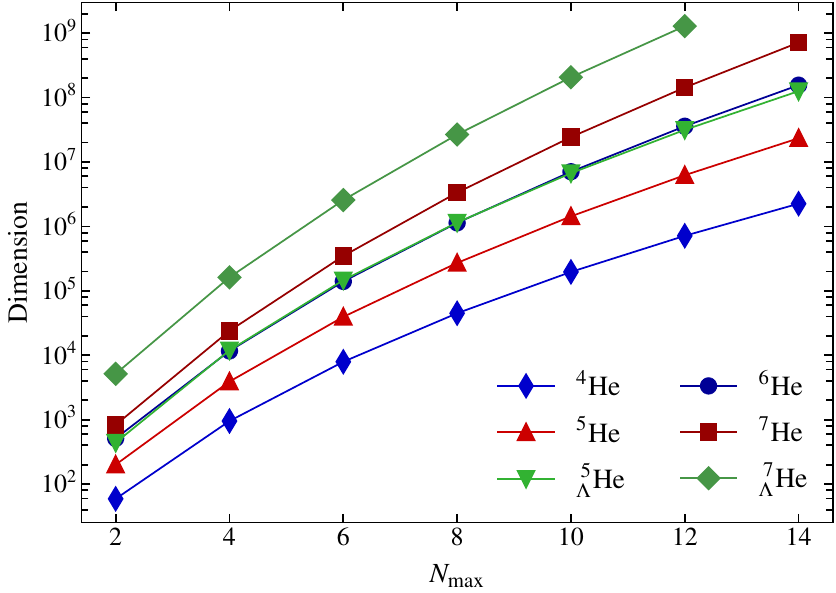}
  \caption{\label{fig:itncsm model space size}(Color online) Dimension of the full NCSM model space for natural parity and $M_J=\{0,1/2\}$ as function of $\Nmax$ for the two hypernuclei \isotope[5][\Lambda]{He} and \isotope[7][\Lambda]{He}, compared to their nucleonic parents and the nuclei obtained by replacing the hyperon with a neutron.}
\end{figure}

The model space grows rapidly as a function of $\Nmax$.
Since the inclusion of the additional particle species further aggravates this behavior, model-space sizes become intractably large even for moderate particle numbers.
As an example, we show the model-space dimensions of \isotope[5][\Lambda]{He} and \isotope[7][\Lambda]{He} in \cref{fig:itncsm model space size}, along with their nucleonic parents and a nucleus with the same particle number $A$.
The model spaces for the hypernuclei are two orders of magnitude larger than those for the parent nuclei and increase at about the same rate as the model spaces for the nuclei.
The dimension for \isotope[7][\Lambda]{He} at $\Nmax=12$ is \num[scientific-notation=true,round-mode=places,round-precision=1]{1284704392}, which is at the limit of what current supercomputers can handle when employing three-body forces \cite{Barrett2013}.

However, especially in spaces corresponding to large $\Nmax$, there are many basis states that are irrelevant for the description of low-lying eigenstates of the Hamiltonian.
In an expansion of any of these eigenstates
\begin{equation}
  \ket{\psi_i}=\sum_{k\in\modelspace} C_k^{(i)} \ket{\phi_k}
\end{equation}
the expansion coefficients $C_k^{(i)}$ of many basis states have very small values.
We can, therefore, exclude these states from the model space, dramatically reducing the dimension of the problem, while affecting the states we wish to describe only to a small degree.
In order to identify the irrelevant basis states without actually solving the eigenvalue problem in the full space $\modelspace$, we need an \emph{a priori} estimate of the coefficient $C_k^{(i)}$ of a basis state $\ket{\phi_k}$ in the expansion of $\ket{\psi_i}$.

We adapt the approach employed in Ref.~\cite{Roth2009} to get this estimate:
we consider the first-order perturbative correction
\begin{equation}
    \ket{\psi_i^{(1)}} = -\sum_{\mathclap{k\notin\modelspace^\text{ref}}}  \frac{\braket{\phi_k|\op{H}|\psi_i^\text{ref}}}{\epsilon_k - \epsilon_\text{ref}}\ket{\phi_k}
\end{equation}
to a reference state $\ket{\psi_i^\text{ref}}$ from a small model space $\modelspace^\text{ref}$.
This state should approximate the target state to be described and may be obtained, e.g., from a diagonalization of the Hamiltonian in $\modelspace^\text{ref}$.
The energy $\epsilon_\text{ref}$ is either the associated eigenvalue or the sum of single-particle energies.

The unperturbed energy $\epsilon_k$ defines the partitioning of the Hamiltonian and can be chosen freely.
We use a simple M\o{}ller--Plesset-type choice
\begin{equation}
  \epsilon_k = \epsilon_\text{ref} + \Delta\epsilon_k + \Delta M_k,
\end{equation}
where, just like in Ref.~\cite{Roth2009}, $\Delta\epsilon_k$ is the excitation energy of $\ket{\phi_k}$ above the HO ground state.
The term $\Delta M_k$ is added for the hypernuclear case and accounts for the difference between the total masses of all particles in  $\ket{\phi_k}$ and in the HO ground state, which is built with $\Lambda$ hyperons.
In the case of single-hyperon systems $\Delta M_k$ is approximately equal to the mass difference between the $\Lambda$ and the $\Sigma^{0,\pm}$ if $\ket{\phi_k}$ contains a $\Sigma^{0,\pm}$ hyperon and zero otherwise.
For the charged $\Sigma$ hyperons an additional small contribution to $\Delta M_k$ arises from the simultaneous conversion of a proton to a neutron (or vice versa).

With these unperturbed energies, we define the importance measure
\begin{equation}
  \kappa_k^{(i)} = -\frac{\braket{\phi_k|\op{H}|\psi_i^\text{ref}}}{\epsilon_k - \epsilon_\text{ref}} = -\sum_{\mathclap{m\in\modelspace^\text{ref}}} C_m^{(i,\text{ref})} \frac{\braket{\phi_k|\op{H}|\phi_m}}{\Delta\epsilon_k+\Delta M_k},
  \label{eq:itncsm:importance measure}
\end{equation}
where the $C_m^{(i,\text{ref})}$ are the expansion coefficients of the $i$th reference state.
We construct an importance-truncated model space $\modelspace^\text{IT}$ by including only those states $\ket{\phi_k}$ with
\begin{equation}
  \lvert\kappa_k^{(i)}\rvert \geq \kappa_\text{min}
  \label{eq:itncsm:kmin criterion}
\end{equation}
for any reference state~$i$.
The resulting space is tailored to the description of the given set of reference states.
However, the basis states discarded by the criterion \eqref{eq:itncsm:kmin criterion} still have an effect on observables.
We account for this effect by extrapolating to vanishing $\kappa_\text{min}$ threshold using the same extrapolation procedure as described in Ref.~\cite{Roth2009}.

In the limit of vanishing threshold $\kappa_\text{min}$ the model-space construction procedure creates the full $\Nmax=N+2$ space if the reference state is from an $\Nmax=N$ space.
This provides the foundation for an iterative procedure:
We start by performing full NCSM calculations up to a morderate value of $\Nmax$, e.g., $\Nmax=6$ for the lower $p$ shell.
Next, we take the ground and a few excited states from the $\Nmax=6$ calculation, build the importance-truncated $\Nmax=8$ model space, and diagonalize the Hamilton matrix for a sequence of $\kappa_\text{min}$ values using the Lanczos algorithm \cite{Lanczos1950,Lehoucq1997}.
This yields sequences of eigenenergies and eigenstates from which we can compute other observables like radii or transition strengths.
The $\kappa_\text{min}$ sequence is necessary for the threshold extrapolation and can be calculated efficiently by constructing the model space for the lowest value of the sequence and then successively raising the threshold, removing rows and columns from the Hamilton matrix that correspond to basis states falling below the new threshold.
The eigenstates obtained in the model space with the lowest threshold are then used as reference states for the construction of the $\Nmax=10$ model space and the procedure is repeated.
Since the model-space construction considers particle-hole excitations on each of the reference-space basis states, the importance of each basis state of $\modelspace^\text{IT}$ is reassessed in every iteration using the updated approximations to the target states.

The calculation of the importance measure is the most time-consuming part of the model-space construction.
By reducing the number of basis states in the reference space $\modelspace^\text{ref}$ we can speed up this part of the calculation because less matrix elements have to be calculated.
Therefore, we introduce a coefficient threshold $C_\text{min}$ and keep only those basis states from the reference space whose coefficients $C_m^{(i,\text{ref})}$ in the expansion of the target states $\ket{\psi_i}$ exceed this threshold, i.e.,
\begin{equation}
  \lvert C_m^{(i,\text{ref})}\rvert \ge C_\text{min}
\end{equation}
for any of the target states.
If we keep the threshold $C_\text{min}$ sufficiently low, the change of the importance-truncated model space created this way is minimal because many states of the full model space $\modelspace$ remain reachable by excitation of more than one parent state from the reference-space basis.

\begin{figure}
  \centering
  \includegraphics{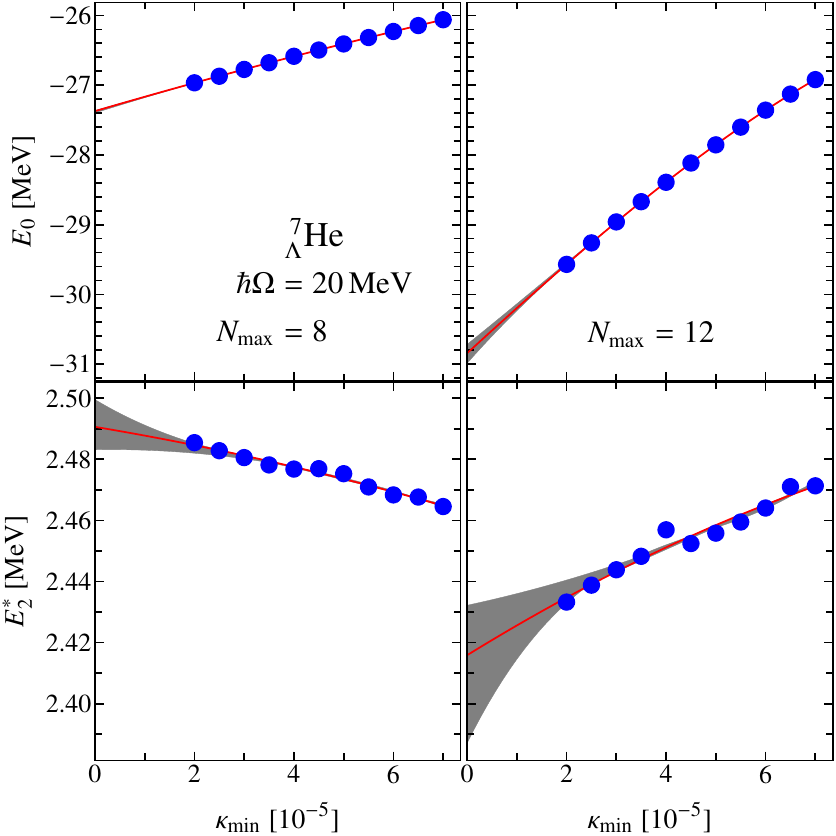}
  \caption{\label{fig:itncsm:extrapolation}(Color online) Threshold extrapolation of the absolute ground-state energy and the excitation energy of the second excited state in \isotope[7][\Lambda]{He}.
  The computed values (blue dots) are shown together with the best-fit polynomial (red line) and the uncertainty due to the unknown functional form, approximated by a set of different fit polynomials (gray band, see text for details).
  The regulator cutoff of the \YN{} interaction is $\Lambda=\SI{700}{\MeVc}$.
  }
\end{figure}

\subsection{Extrapolation Procedure}
The functional form of the value of an observable as a function of the threshold $\kappa_\text{min}$ is unknown.
However, the computed values mostly show a smooth variation with $\kappa_\text{min}$, with tiny variations around the general trend due to the discretization artifacts from the finite model space.
Thus, in order to reliably extract the $\kappa_\text{min}\to0$ value of an observable we perform the extrapolation by fitting low-order polynomials to the computed values.
We take a cubic polynomial to extract the best-fit extrapolated value for absolute energies; the extrapolation uncertainties are estimated by fitting a quadratic and quartic polynomial to the same data set, and by fitting cubic polynomials with the lowest threshold or the lowest two thresholds excluded from the fit.
Analogously, excitation energies are extrapolated with quadratic, linear, and cubic polynomials.
The resulting best-fit polynomial and error bands are shown in \cref{fig:itncsm:extrapolation} for the ground-state energy and the energy of the second excited state in \isotope[7][\Lambda]{He}.
Note that although we have ${\sim}\SI{100}{\keV}$ uncertainty in the determination of the absolute energy at $\Nmax=12$ the excitation energy can be determined with high precision.

\subsection{\label{sec:itncsm:center of mass}Center of Mass Excitations}
In the $\Nmax$ truncation scheme, the intrinsic wave function of the nucleus decouples from the center-of-mass component.
Hence, the NCSM does not suffer from contamination of the intrinsic state by a coupling to the motion of the center of mass.
However, since the center-of-mass degree of freedom is included in the basis, the spectrum of the intrinsic, translationally invariant Hamiltonian contains states that are formed from identical intrinsic states combined with different excitations of the center-of-mass component.
In the full Hilbert space these states become degenerate, but in the finite $\Nmax$\nobreakdash-truncated model space an excitation of the center-of-mass component takes up HO quanta that are not available for the intrinsic wave function.
Thus, states with center-of-mass excitations have intrinsic wave functions that correspond to lower $\Nmax$ values; they contain no additional information but increase the computational demands of the calculation.

To address this, we add a center-of-mass Hamiltonian \cite{Gloeckner1974} (see \cref{app:tint and com}), which contains a HO potential acting on the center-of-mass coordinate, to the intrinsic Hamiltonian of the system.
The total Hamiltonian that is used in the calculation thus becomes
\begin{equation}
  \Ham = \Ham_\text{int} + \beta_\text{c.m.}\Ham_\text{c.m.},
\end{equation}
where the parameter $\beta_\text{c.m.}$ controls the strength of the center-of-mass Hamiltonian.
In this way, we shift states with a center-of-mass wave function different from the HO ground state to higher energies and remove them from the relevant part of the spectrum.
Due to the decoupling property, the intrinsic wave function is not affected by this procedure.

\section{\label{sec:validation}Validation}
Since the J-NCSM and the IT\nobreakdash-NCSM use the same model spaces, their results not only agree in the infinite-model-space limit but also for each value of \Nmax.
We use this property to validate both implementations in the four-body system.

The Hamiltonian we employ for this consists of interactions derived from chiral effective field theory.
We use a two-nucleon interaction at next-to-next-to-next-to-leading order (N\textsuperscript3LO) \cite{Entem2003}, a three-nucleon interaction at N\textsuperscript2LO \cite{Navratil2007}, both with a regulator cutoff $\Lambda=\SI{500}{\MeVc}$.
The hyperon-nucleon interaction is taken at leading order \cite{Polinder2006}.
From this Hamiltonian we derive two additional ones by subjecting the nucleonic part or the nucleonic and hyperon parts to an SRG transformation.
These will be called ``bare'', ``SRG(3)$_{N}$'' and ``SRG(3)$_{N}$+SRG(2)$_{Y}$'' in the following, where the number in parentheses denotes the particle rank of the initial Hamiltonian and of the included induced many-body terms.

\begin{figure}
  \centering
  \includegraphics{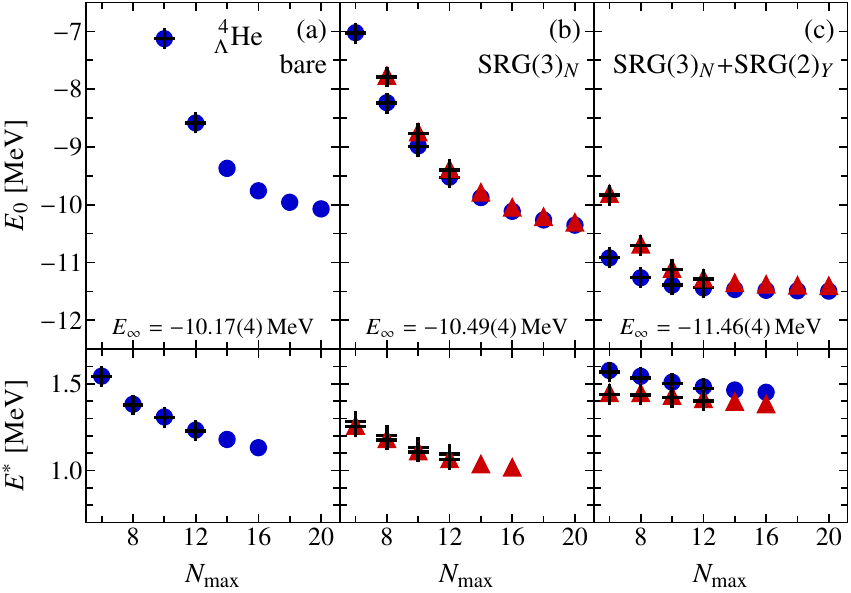}
  \caption{\label{fig:validation:lhe4}%
    (Color online) Ground-state and excitation energy of \isotope[4][\Lambda]{He} for the (a) bare, (b) SRG(3)$_{N}$ evolved to $\alpha=\SI{0.04}{\fm\tothe4}$ (red triangles) and $\alpha=\SI{0.08}{\fm\tothe4}$ (blue circles), and (c) SRG(3)$_{N}$+SRG(2)$_{Y}$ Hamiltonian, obtained from a J-NCSM calculation.
    The regulator cutoff is $\Lambda=\SI{600}{\MeVc}$, the basis frequency is $\hbar\Omega=\SI{20}{\MeV}$ for the evolved Hamiltonians and $\hbar\Omega=\SI{30}{\MeV}$ for the bare one.
    The crosses denote the IT-NCSM results for both flow parameters, $E_\infty$ is the energy extrapolated to $\Nmax\to\infty$.}
\end{figure}

\begin{table}
  \sisetup{table-format=-2.2(2)}
  \begin{tabular}{lSSSS}
    \doubletoprule
                    & \multicolumn{4}{c}{$\Lambda$ [\si{\MeVc}]} \\
    \cmidrule{2-5}
    \multicolumn{1}{c}{Nucleus} & {550} & {600} & {650} & {700} \\
    \midrule
    \isotope[4][\Lambda]{He}($0^+$) & -10.40(6)  & -10.17(4) & -10.07(5) & -10.11(4) \\
    \isotope[4][\Lambda]{He}($1^+$) & -9.47(25)  & -9.06(28) & -8.75(34) & -8.40(30) \\
    \isotope[4][\Lambda]{H}($0^+$)  & -11.20(6)  & -10.95(4) & -10.83(5) & -10.85(3) \\
    \isotope[4][\Lambda]{H}($1^+$)  & -10.29(25) & -9.87(28) & -9.55(35) & -9.20(30) \\
    \doublebottomrule
  \end{tabular}
  \caption{\label{tab:validation:binding energies}%
  Extrapolated ground- and excited-state energies of the $A=4$ systems for different values of the interaction regulator $\Lambda$ using the bare Hamiltonian.
  The ground-state energies of \isotope[3]{He} and \isotope[3]{H} obtained with this Hamiltonian are \SI{-7.72}{\MeV} and \SI{-8.47}{\MeV}, respectively.}
\end{table}

\Cref{fig:validation:lhe4} shows the ground-state and excitation energy of \isotope[4][\Lambda]{He} for the bare, SRG(3)$_{N}$, and SRG(3)$_{N}$+SRG(2)$_{Y}$ Hamiltonian.
In all cases we see excellent agreement between the IT\nobreakdash-NCSM and J\nobreakdash-NCSM formulations down to the level of \SI{10}{\keV}.
This difference stems from the different treatment of isospin in both approaches:
The IT-NCSM uses a particle basis throughout the calculation, while the J-NCSM transforms the particle-basis matrix elements into an isospin-coupled basis (cf.\ \cref{app:pbasisintbasis}).

For the bare interaction, we see slow convergence of the energies, the last step in $\Nmax$ still lowering the ground-state energy by \SI{114}{\keV}.
The SRG(3)$_{N}$ Hamiltonian improves convergence for smaller $\Nmax$ values; at larger model spaces, the convergence behavior is unchanged and dominated by the unevolved \YN{} interaction.
This improved convergence comes at the price of repulsive induced many-body terms of about \SI{300}{\keV}.
Since there are only three nucleons, these terms are \YNN{} (and possibly \YNNN{}) interactions.

The SRG(3)$_{N}$+SRG(2)$_{Y}$ Hamiltonian drastically improves convergence of both the ground-state and excitation energy.
The ground-state energy is now converged to better than \SI{10}{\keV}.
However, we see from the difference between the extrapolated binding energies that the evolution of the \YN{} interaction induces sizable repulsive many-body terms of \SI{1.3}{\MeV}.
Moreover, the excitation energy predicted by the SRG(3)$_{N}$+SRG(2)$_{Y}$ Hamiltonian differs significantly from the predictions by the other Hamiltonians.
These effects become stronger with larger systems and are the reason why we used bare \YN{} interactions in Ref.\ \cite{Wirth2014}.

\Cref{tab:validation:binding energies} shows the extrapolated absolute binding energies of the ground and first excited states of \isotope[4][\Lambda]{He} and \isotope[4][\Lambda]{H}.
The resulting $\Lambda$ separation energies agree within uncertainties with those found in \cite[Table 5]{Haidenbauer2007}.

\section{\label{sec:hyperhelium}Helium Hypernuclei}
\begin{figure}
  \centering
  \includegraphics{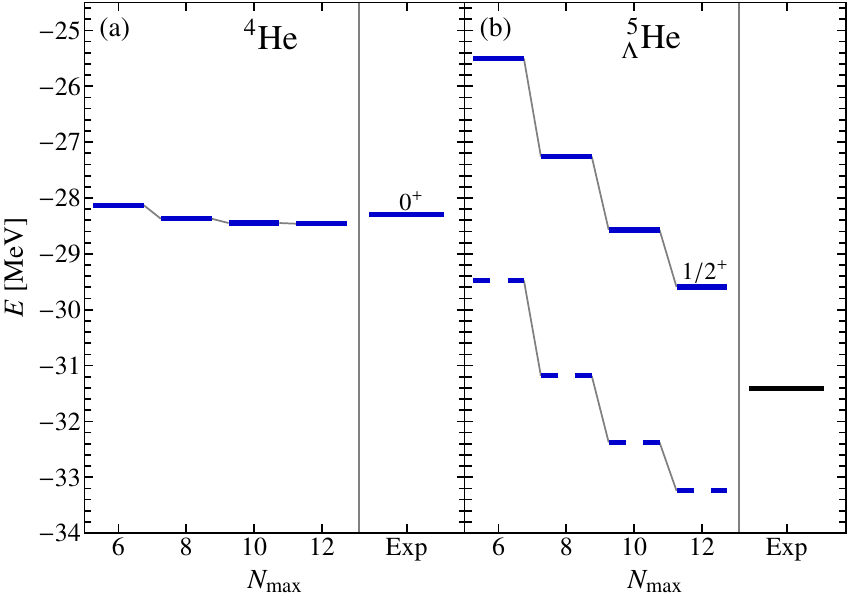}
  \caption{\label{fig:validation:lhe5}%
    (Color online) Absolute ground-state energy of (a) \isotope[4]{He} and (b) its daughter hypernucleus \isotope[5][\Lambda]{He}.
    The oscillator frequency is $\hbar\Omega=\SI{20}{\MeV}$.
    We use the SRG(3)$_{N}$ Hamiltonian with a flow parameter of $\alpha_N=\SI{0.08}{\fm\tothe4}$ in the nucleonic sector and no SRG evolution, i.e., $\alpha_Y=\SI{0}{\fm\tothe4}$, of the YN interaction.
    The regulator cutoffs employed are $\Lambda=\SI{700}{\MeVc}$ (solid lines) and $\Lambda=\SI{600}{\MeVc}$ (dashed lines).
    Experimental data from Refs.~\cite{Wang2012,Davis2005}.
  }
\end{figure}

\begin{figure}
  \centering
  \includegraphics{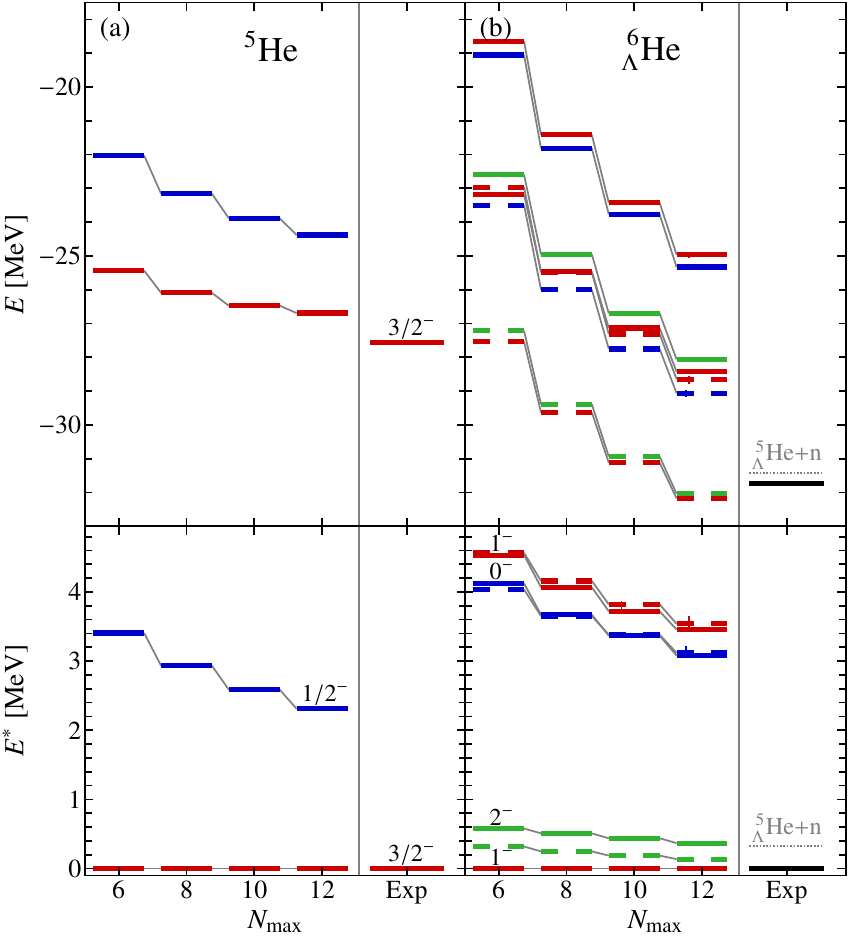}
  \caption{\label{fig:validation:lhe6}%
    (Color online) Absolute and excitation energies of low-lying states in (a) \isotope[5]{He} and (b) its daughter hypernucleus \isotope[6][\Lambda]{He}.
    Parameters are the same as in \cref{fig:validation:lhe5}.
    Vertical lines mark extrapolation uncertainties, experimental data taken from Refs.~\cite{Wang2012,Davis2005}.
  }
\end{figure}

\begin{figure}
  \centering
  \includegraphics{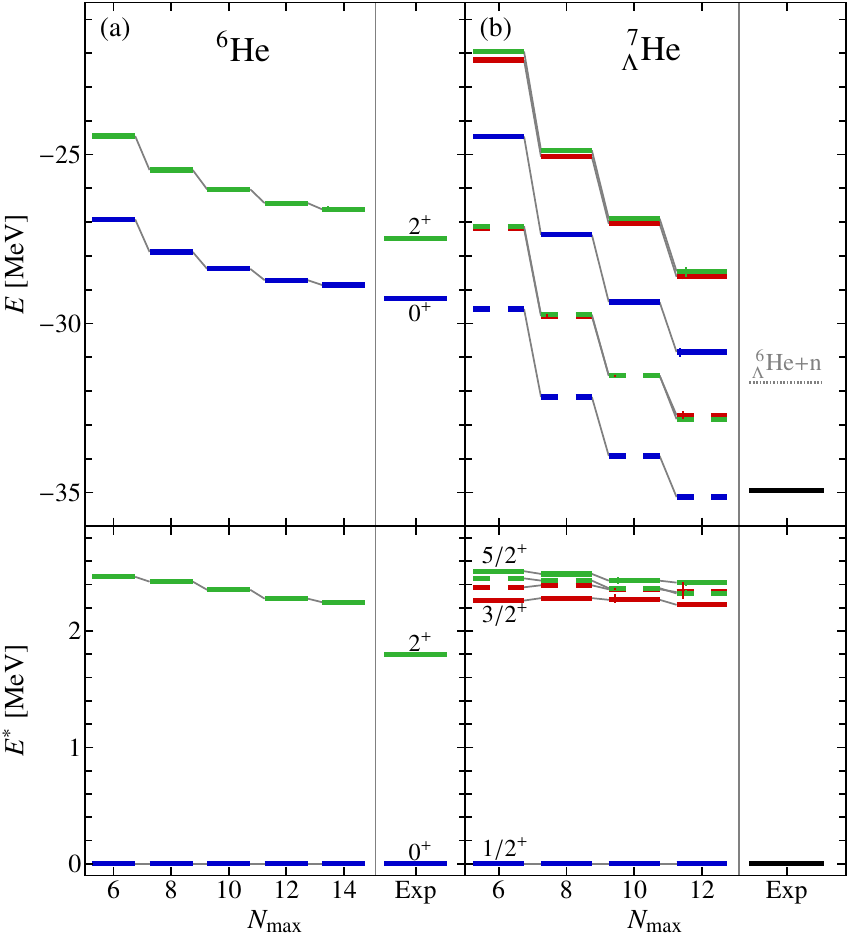}
  \caption{\label{fig:validation:lhe7}%
    (Color online) Absolute and excitation energies of low-lying states in (a) \isotope[6]{He} and (b) its daughter hypernucleus \isotope[7][\Lambda]{He}.
    Parameters are the same as in \cref{fig:validation:lhe5}.
    Vertical lines mark extrapolation uncertainties, experimental data taken from Refs.~\cite{Wang2012,Nakamura2013}.
  }
\end{figure}

Going beyond the four-body system, we explore the helium hypernuclei \isotope[5,6,7][\Lambda]{He} in \cref{fig:validation:lhe5,fig:validation:lhe6,fig:validation:lhe7}, using the SRG(3)$_{N}$ Hamiltonian.
Just like in the four-body system, convergence is dominated by the unevolved \YN{} interaction and the absolute energies are far from convergence.
To extract precise separation energies for these systems an SRG evolution in the \YN{} sector is essential, which has to be done in three-body space to account for the strong induced \YNN{} terms.
Overall, the absolute energies show a strong dependence on the regulator cutoff where the $\Lambda=\SI{600}{\MeVc}$ interaction consistently provides about \SIrange{4}{5}{\MeV} more binding; a trend that is much weaker in the $A=4$ systems.

Energy differences show better convergence.
The excited state in the \isotope[5]{He} spectrum (\cref{fig:validation:lhe6}) is a broad resonance that is hard to describe in a HO basis and, thus, converges slowly.
The experimental data show that the additional attraction due to the hyperon in the daughter hypernucleus \isotope[6][\Lambda]{He} barely stabilizes the ground state against neutron emission.
All excited states, except for the $2^-$, are well in the \isotope[5][\Lambda]{He}+n continuum.
The excitation energy of the excited-state doublet exhibits a convergence pattern that is similar to the $1/2^-$ in \isotope[5]{He}, but qualitatively different from the pattern exhibited by the bound excited states in \isotope[7][\Lambda]{He}.
This hints at these states being broad resonances.
The ground-state doublet splitting is sensitive to the \YN{} regulator cutoff, while the excited-state doublet shows only little variation.

The excited-state doublet in \isotope[7][\Lambda]{He} (\cref{fig:validation:lhe7}) shows convergence on par with the nucleonic parent.
Like the doublet in \isotope[9][\Lambda]{Be} it is almost degenerate (cf.\ Ref.~\cite{Wirth2014}), and the $\Lambda=\SI{600}{\MeVc}$ interaction predicts the $5/2^+$ to be the lower state of the doublet, while the higher cutoff shows a relatively constant doublet splitting with the $3/2^+$ as lower state.
In a microscopic shell-model calculation \cite{Millener2007} the \isotope[6]{Li} analog state of the $2^+$ excitation in \isotope[6]{He} is dominantly an $L=2$, $S=0$ state so that our results confirm that the chiral interaction correctly reproduces the smallness of the effective hyperon-nucleus spin-orbit interaction that is observed in these calculations.

The admixture of $\Sigma$ hyperons is at the level of \numrange{1.6e-2}{1.9e-2}.
There are differences of the order of $\num{3e-4}$ among the doublet partners, as well as between the two cutoffs.
This indicates that the $\Lambda N$-$\Sigma N$ conversion has a small contribution to the doublet splittings.
However, the differences in binding energies between the two cutoffs cannot be explained by the small change in the admixture and, thus, are likely caused by a different part of the \YN{} interaction.

Overall, we see that while absolute energies are far from convergence, except for the $A=4$ systems, the excitation energies show much less variation.
Already at this level, we can use these results to investigate properties of the different \YN{} interactions, e.g., the cutoff dependence of the LO potential used here, or to confront them with experimental data in order to select interaction parameters based on how well the data are reproduced.

\section{Conclusions and Outlook}
We present an extension of the NCSM for the \emph{ab initio} description of single\nobreakdash-$\Lambda$ hypernuclei.
Starting from the general form of the hypernuclear Hamiltonian with \NN{}, \YN{}, and \NNN{} interactions we show the SRG transformation of the two-baryon interactions at the two-body level, which improves convergence of the many-body method by evolving the Hamiltonian towards diagonal form in momentum space.
We continue with the NCSM itself and provide two complementary formulations, each with their own merits and drawbacks: a Jacobi-coordinate formulation (J-NCSM) that fully exploits the rotational, translational, and isospin symmetries of the Hamiltonian but requires explicit antisymmetrization, which becomes computationally more and more demanding with increasing particle number, and a formulation in single-particle coordinates where antisymmetry is implemented via a Slater Determinant basis, but where center-of-mass degrees of freedom are included in the model space and only a subset of the symmetries can be exploited.
The single-particle formulation allows for a selection of relevant basis states via an importance truncation, leading to the IT-NCSM.

We demonstrate that both formulations give the same results in the $A=4$ system, and present a compilation of absolute ground- and excited-state energies for \isotope[4][\Lambda]{H} and \isotope[4][\Lambda]{He} for the \YN{} interaction from leading-order chiral EFT with a state-of-the-art chiral nucleonic Hamiltonian containing \NN{} and \NNN{} interactions.
We also show that, even in the four-body system, the SRG evolution of the \YN{} interaction induces strong repulsive \YNN{} terms that have to be included explicitly in order to get precise results.
Finally, we conclude with a survey of the hyper-helium chain.

The hypernuclear NCSM enables precise \emph{ab initio} calculations of hypernuclear observables for few-body systems and, with inclusion of SRG-induced \YNN{} terms done in \cite{Wirth2016}, throughout the $p$ shell.
It thus provides a valuable link between the \YN{} interaction and the wealth of experimental data available on light hypernuclei, especially considering the scarcity of \YN{} scattering data.
We can use this link to tune interaction parameters toward a good reproduction of the available data on some of the light hypernuclei in order to make predictions for others.
This more phenomenological approach might be necessary because the available scattering data are barely constraining the five low-energy constants of the leading-order interaction.

Furthermore, the hypernuclear NCSM provides a testing ground for validating other methods that are suitable for medium-mass systems, making it the first step towards an \emph{ab initio} many-body theory for these hypernuclei.
We can also benchmark approaches like the Brueckner-Hartree-Fock method that makes hypernuclei with closed-shell parents easily accessible throughout the hypernuclear chart, and can also be used to compute the equation of state of hypernuclear matter \cite{Vidana2011,Petschauer2015}.

With this framework, we are also able to address interesting physics questions in the $p$ shell, e.g., whether the attraction provided by the hyperon can shift the neutron drip line \cite{LightNeutronRich}.
The high precision of the J-NCSM can be used, e.g., to study the charge-symmetry breaking in the $A=4$ system \cite{Gazda2016a,*Gazda2016b}.

% "I wrote the book on the appendix. I even wrote the appendix, but they took that out." -- Cpt. B.F. Pierce, 4077th MASH
\appendix

\section{\label{app:tint and com}Intrinsic Kinetic Energy and Center-of-Mass Hamiltonian}
The presence of particles with different masses changes the way operators acting on intrinsic or center-of-mass coordinates are calculated compared to the nucleonic case.
This appendix outlines the derivation of expressions for the intrinsic kinetic energy and the center-of-mass Hamiltonian.
Expressions for other operators like mean-square radii can be obtained in an analogous way.

The intrinsic kinetic energy $\Tint$ is calculated in the following way:
we express the total mass and the center-of-mass coordinate and momentum,
\begin{equation}
  M =\sum_{i=1}^A m_i,\quad \vect{R} = \frac{1}{M}\sum_{i=1}^A m_i\vect{r}_i, \quad \vect{P} = \sum_{i=1}^A \vect{p}_i
\end{equation}
in terms of the absolute single-particle coordinates $\vect{r}_i$, momenta $\vect{p}_i$ and masses $m_i$.
The coordinates $\vect{\rho}_i$ and momenta $\vect{\pi}_i$ of the particles relative to the center of mass then become
\begin{align}
  \vect{\rho}_i &= \vect{r}_i - \vect{R} = \vect{r}_i - \frac{1}{M}\sum_{j=1}^A m_j\vect{r}_j \\
  \vect{\pi}_i &= \vect{p}_i - \frac{m_i}{M}\sum_{j=1}^A \vect{p}_j = \vect{p}_i - \frac{m_i}{M}\vect{P}.
\end{align}
Additionally, we have, for each pair of particles $i$ and $j$, the relative coordinates and momenta in their center-of-mass system:
\begin{align}
  \vect{r}_{ij} &= \vect{r}_j - \vect{r}_i \\
  \vect{q}_{ij} &= \mu_{ij} \biggl(\frac{\vect{p}_j}{m_j} - \frac{\vect{p}_i}{m_i}\biggr),
\end{align}
where $\mu_{ij} = m_im_j/(m_i+m_j)$ is the reduced mass of particles $i$ and $j$.
Using these coordinates, we write the intrinsic kinetic energy as
\begin{align}
  T_\text{int} &= \sum_{i=1}^A \frac{\pi_i^2}{2m_i}
    = \sum_{i=1}^A \frac{1}{2m_i} \biggl(\vect{p}_i - \frac{m_i}{M}\sum_{j=1}^A \vect{p}_j\biggr)^2 \notag\\
   &= \sum_{i=1}^A \frac{p_i^2}{2m_i} -\frac{1}{2M}\sum_{i,j=1}^A m_im_j\frac{\vect{p}_i\cdot\vect{p}_j}{m_im_j},
  \label{eq:tint and com:tint step1}
\end{align}
where we used the definition of the total mass $M$ to eliminate sums over particle masses.
The scalar product can be expressed in terms of squared momenta:
\begin{equation}
  \frac{\vect{p}_i\cdot\vect{p}_j}{m_im_j} = \frac{p_i^2}{2m_i^2} + \frac{p_j^2}{2m_j^2} - \frac{q_{ij}^2}{2\mu_{ij}^2}.
  \label{eq:tint and com:pipj}
\end{equation}
Inserting this into \cref{eq:tint and com:tint step1} and simplifying, we get
\begin{align}
  T_\text{int}
   &= \frac12 \sum_{ij}\frac{m_i+m_j}{M} \frac{q_{ij}^2}{2\mu_{ij}} = \sum_{i<j}\frac{m_i+m_j}{M} T_{ij,\text{rel}},
\end{align}
where $T_{ij,\text{rel}} = q_{ij}^2/(2\mu_{ij})$ is the relative kinetic energy in the two-body system of particles $i$ and $j$.
Formally, this operator is an $A$\nobreakdash-body operator due to the $1/M$ factor but it can be evaluated using the Slater rules for a two-body operator after exploiting that the basis determinants are eigenstates to the total mass operator $\Mass$.

As discussed in \cref{sec:itncsm} we lift the degeneracy of eigenstates of the intrinsic Hamiltonian with respect to the center-of-mass state by adding a harmonic-oscillator Hamiltonian
\begin{equation}
  \Ham_\text{c.m.} = \frac{1}{2M}\vect{\op{P}}^2 + \frac12 M \Omega^2 \vect{\op{R}}^2 -\frac32\Omega
\end{equation}
acting on the center-of-mass coordinate $\vect{R}$ and momentum $\vect{P}$.
The offset is introduced so that the ground state has zero energy.
Using relation \cref{eq:tint and com:pipj} for the single-particle momenta $\vect{p}_i,\vect{p}_j$, this Hamiltonian separates into zero-, one- and two-body parts:
\begin{align}
  \MoveEqLeft\frac{1}{2M}P^2 + \frac12 M \Omega^2 R^2 -\frac32 \Omega \notag\\*
  &=\frac{1}{2M}\sum_{i=1}^A\sum_{j=1}^A\bigl(\vect{p}_i\cdot\vect{p}_j + \Omega^2 m_im_j \vect{r}_i\cdot\vect{r}_j\bigr)-\frac32\Omega \notag\\
  &=\sum_{i} \biggl(\frac{p_i^2}{2m_i} + \frac12 m_i \Omega^2 r_i^2\biggr) \notag\\
  &\pheq -\sum_{i<j}\frac{m_i+m_j}{M} \biggl(\frac{q_{ij}^2}{2\mu_{ij}} + \frac12 \mu_{ij} \Omega^2 r_{ij}^2\biggr)-\frac32\Omega.
\end{align}
Hence, the center-of-mass Hamiltonian separates into a constant part, a harmonic-oscillator Hamiltonian for each of the particles and a relative two-body hamonic oscillator for each particle pair.
The one-body part is evaluated by using an eigenvalue relation with respect to the single-particle states; the two-body part may be calculated using Slater rules after using the eigenvalue relation of $\Mass$.

\section{Evaluation of particle-basis defined potentials in isospin basis}
\label{app:pbasisintbasis}
Realistic \NN{} and \NY{} interactions are usually defined in the
particle basis, not in the isospin basis. To evaluate the corresponding matrix
elements of $\op{V}_{\NN}$ between good-isospin basis states
\eqref{eq:jncsm:brho} we use
the following prescription
\begin{align}
\MoveEqLeft\braket{((T_{\Am3}',t')T_{\Am1}',t_{\Am1}')TM_T| V_{\NN}
|(T_{\Am3},t)T_{\Am1},t_{\Am1}) TM_T}
\notag\\ &=
\kronecker{T_{\Am3}'}{T_{\Am3}}
\kronecker{t'}{t}
\kronecker{T_{\Am1} '}{T_{\Am1}}
\kronecker{t_{\Am1}'}{t_{\Am1}}
\sum
\notag\\
&\pheq\times \left(\clebsch{T_{\Am1}}{M_{\Am1}}{t_{\Am1}}{m_{\Am1}}{T}{M_T}\right)^2
\left(\clebsch{T_{\Am3}}{M_{\Am3}}{t}{m}{T_{\Am1}}{M_{\Am1}}\right)^2\notag\\
&\pheq\times \clebsch{\tfrac{1}{2}}{m_a'}{\tfrac{1}{2}}{m_b'}{t}{m}
\clebsch{\tfrac{1}{2}}{m_a}{\tfrac{1}{2}}{m_b}{t}{m} \notag\\
&\pheq\times \braket{\tfrac{1}{2}\, m_a',\tfrac{1}{2}\,m_b' | V_{\NN} |
\tfrac{1}{2}\, m_a,\tfrac{1}{2}\,m_b} \notag\\
&\equiv V_{\NN}(t;T_{\Am3},T_{\Am1},t_{\Am1},T,M_T). \label{eq:jncsm:vnnavg}
\end{align}
Here, only the isospin quantum numbers of the states \eqref{eq:jncsm:brho} are displayed.
The states are decomposed via Clebsch--Gordan coefficients and the
potential matrix elements are evaluated between two-nucleon states $\ket{
\tfrac{1}{2}\,m_a,\tfrac{1}{2}\,m_b}$ with single-nucleon isospin
projections $m_a=\pm\frac{1}{2}$ and $m_b=\pm\frac{1}{2}$. In this procedure
the isospin breaking transitions $t=0 \leftrightarrow 1$ are suppressed.

Similarly, a particle-basis defined \NY{} interaction $V_{\NY}$
is evaluated in basis \eqref{eq:jncsm:beta} as
\begin{align}
\MoveEqLeft\braket{ (T_{\Am2}',t'\tau')TM_T | V_{\NY} | (T_{\Am2},t\tau)TM_T }
\notag\\&=\kronecker{t_{\Am2}'}{t_{\Am2}} \kronecker{t'}{t}
\sum \left(\clebsch{T_{\Am2}}{M_{\Am2}}{t}{m}{T}{M_T}\right)^2\notag\\
&\pheq\times
\clebsch{\tau'}{m_\tau'}{\tfrac{1}{2}}{m_b'}{t}{m}
\clebsch{\tau}{m_\tau}{\tfrac{1}{2}}{m_b}{t}{m} \notag\\
&\pheq\times \braket{\tau' \,m_\tau', \tfrac{1}{2}, m_b' | V_{\NY} |
\tau \,m_\tau, \tfrac{1}{2}\, m_b} \notag\\
&\equiv V_{\NY}(t,\tau',\tau;T_{\Am2},T,M_T), \label{eq:jncsm:vnyavg}
\end{align}
where the potential matrix elements are evaluated between hyperon--nucleon
states $\ket{\tau\, m_\tau, \tfrac{1}{2}\, m_b}$ with $m_\tau=-\tau,0,\tau$ and
$m_b=\pm\frac{1}{2}$ the isospin projections of $\Lambda$ ($\tau=0$) or $\Sigma$
($\tau=1$) hyperon and nucleon, respectively. Again, the isospin-breaking
transitions $t=\frac{1}{2} \leftrightarrow\frac{3}{2}$ are suppressed.

The procedure described above gives excellent agreement with particle-basis
calculations as demonstrated in Section \ref{sec:validation}. For the $A$=3,4
hypernuclear systems, the difference between calculated total energies in
particle basis and isospin basis using relations \eqref{eq:jncsm:vnnavg} and
\eqref{eq:jncsm:vnyavg} is found to be only few \si{\keV}.

\section*{Acknowledgments}
The research of D.G.\ was supported by the Grant Agency of the Czech Republic (GACR), grant P203/15/04301S.
P.N.~acknowledges support by the NERSC Grant No.\ SAPIN-2016-00033.
TRIUMF receives federal funding via a contribution agreement with the National Research Council of Canada.
R.R.\ and R.W.\ gratefully acknowledge support by Deutsche Forschungsgemeinschaft through SFB 1245, the Helmholtz International Center for FAIR, and the BMBF through Contract No.\ 05P15RDFN1.

Calculations for this research were conducted on the Lichtenberg high performance computer of TU Darmstadt and on the supercomputer JURECA \cite{JURECA} at Forschungszentrum Jülich.

\renewcommand{\doibase}[0]{https://doi.org/}%
\newcommand{\Eprint}[0]{\hfil\penalty100\hfilneg\href }

\end{document}